\documentclass{vgtc}                          

\ifpdf
  \pdfoutput=1\relax                   
  \usepackage{graphicx}                
  \DeclareGraphicsExtensions{.pdf,.png,.jpg,.jpeg} 
\else
  \usepackage{graphicx}                
  \DeclareGraphicsExtensions{.eps}     
\fi%

\graphicspath{{figures/}{pictures/}{images/}{./}} 

\usepackage{microtype}                 
\PassOptionsToPackage{warn}{textcomp}  
\usepackage{textcomp}                  
\usepackage{mathptmx}                  
\usepackage{amssymb}
\usepackage{amsmath}
\usepackage{times}                     
\usepackage{cite}                      
\usepackage{tabu}                      
\usepackage{booktabs}                  
\usepackage{graphicx} 
\usepackage{float} 
\usepackage{subfigure}

\usepackage{multirow}
\usepackage{booktabs}
\usepackage{color}

\newcommand{\zlz}[1]{\textcolor{black}{#1}}

\onlineid{0}

\vgtccategory{Research}

\vgtcinsertpkg

\title{Classifying In-Place Gestures with End-to-End Point Cloud Learning}

\author{
Lizhi Zhao \thanks{Equal Contribution.} \\ \scriptsize Northwest A\&F University
\and Xuequan Lu \footnotemark[1] \\\scriptsize Deakin University
\and Min Zhao \\\scriptsize Northwest A\&F University %
\and Meili Wang  \thanks{Corresponding author: wml@nwsuaf.edu.cn \newline \hspace*{1.8em}Key Laboratory of Agricultural Internet of Things, Ministry of Agriculture, Yangling 712100, China \newline \hspace*{1.5em} Shaanxi Key Laboratory of Agricultural Information Perception and Intelligent Service, Yangling 712100, China} \\ \scriptsize Northwest A\&F University 
}

\abstract{
Walking in place for moving through virtual environments has attracted noticeable attention recently. Recent attempts focused on training a classifier to recognize certain patterns of gestures (e.g., standing, walking, etc) with the use of neural networks like CNN or LSTM.  Nevertheless, they often consider very few types of gestures and/or induce less desired latency in virtual environments. In this paper, we propose a novel framework for accurate and efficient classification of in-place gestures. 
Our key idea is to treat several consecutive frames as a ``point cloud''. The HMD and two VIVE trackers provide three points in each frame, with each point consisting of 12-dimensional features (i.e., three-dimensional position coordinates, velocity, rotation, angular velocity). 
We create a dataset consisting of 9 gesture classes for virtual in-place locomotion. In addition to the supervised point-based network, we also take unsupervised domain adaptation into account due to inter-person variations. To this end, we develop an end-to-end joint framework involving both a supervised loss for supervised point learning and an unsupervised loss for unsupervised domain adaptation. Experiments demonstrate that our approach generates very promising outcomes, in terms of high overall classification accuracy (95.0\%) and real-time performance (192ms latency). \textit{Our code will be publicly available at: https://github.com/ZhaoLizz/PCT-MCD}.
} 

\CCScatlist{
  \CCScatTwelve{Human-centered computing}{Human computer interaction (HCI)}{Interaction techniques}{Gestural input};
  \CCScatTwelve{Human-centered computing}{—Human computer interaction (HCI)}{Interaction paradigms}{Virtual reality}
}

\begin{document}


\firstsection{Introduction}

\maketitle


Locomotion in the virtual environment is an important function in virtual reality (VR) applications \cite{wipservy}. An ideal, immersive and natural approach to virtual locomotion is real walking \cite{10.1145/3013971.3014010}; however the limitation of physical space often makes this approach difficult to implement \cite{williams2007exploring}.  Joystick-based movement is a common method in video games, but this method can cause motion sickness in virtual reality environments \cite{jaeger2001comparison}. The Walking-in-place (WIP) method allows the user to navigate the virtual environment with walking-in-place gestures, which provides kinesthetic feedback similar to real walking and provides high immersion and naturalness \cite{wipservy}. 

Although the WIP technique has the advantages mentioned above, it is still not widely used mainly due to several challenges.  
The first challenge is to classify gesture patterns that map the user's specific gestures in the real world to virtual locomotion with low latency  \cite{felberbaum2018better}. 
There are many proposed WIP methods to recognize specific WIP gestures of users and then to transform them into virtual movements. For instance, LLCM-WIP \cite{LLCMWIP}, GUD-WIP \cite{GUDWIP} and SAS-WIP \cite{bruno2013new} implemented low-latency WIP gestures detection by manually designing features. Hansen \textit{et al.} \cite{hanson2019improving} used a convolutional neural network to classify whether the user is standing or WIP. VR-STEP \cite{tregillus2016vr} utilized a dynamic threshold method proposed 
in \cite{zhao2010full} to detect WIP gait. 
However, many existing methods suffer from manually designing features such as threshold parameters, or introducing high classification latency, or focusing  on classifying only two types of gestures (standing and walking) while ignoring other common in-place gestures (e.g., jogging or jumping).  Moreover, due to the inter-person variations which represent different characteristics of users such as height, weight, gender and locomotion habits, domain gaps do exist between users. So the 
method designed may be accurate for some users, but may not work for others.  
DCTC \cite{shi2019accurate}  is based on improved Long Short-Term Memory (LSTM) network for classifying 7 types of foot patterns and can be  generalized across different sensors and individuals, but it induces a relatively high latency (i.e., 500ms for 95\% accuracy). 

To address the above challenges in virtual locomotion, we propose an effective method that can accurately classify the user's in-place  gestures in the real world with low latency and can bridge the domain gap between users so that the model is generalizable across persons. 
Our key idea is to use the head-mounted display (HMD) and two VIVE trackers attached to the thighs to collect the user's head and legs position, velocity, rotation angle and angular velocity data in each frame, and treat the data of several consecutive frames as a point cloud. Thus each point in the point cloud consists of 12-dimensional features which provide rich geometric, trajectory shape and scale information of user's WIP gestures \cite{guo2013rotational,guo20143d}. We then design an end-to-end point-based classification network that takes both supervised learning and unsupervised domain adaptation into account, with a supervised loss and an unsupervised loss for joint training. 

As for creating the dataset, we first collect 9 classes of locomotion data from 14 participants using the HMD and two VIVE trackers mentioned above. Then, using the sliding window method, the data is partitioned into samples, with each sample representing a point cloud. 
Then we label each sample with the gesture that appears most number of times in that sample.
For a particular user, we first collect the locomotion data of the user, and define the unlabeled data of the user as the target domain and the labeled data in the dataset as the source domain. 
Inspired by Maximum Classifier Discrepancy (MCD) \cite{saito2018maximum} from image processing and Point Cloud Transformer (PCT) \cite{guo2021pct} from point cloud learning, we design an end-to-end framework by fusing MCD with PCT.
With the adversarial training strategy proposed in MCD, 
the encoder from PCT is used as a feature generator, and two task-specific classifiers are used as the discriminator which takes features from the generator as input. 
Two classifiers are trained to simultaneously classify source samples and detect target samples that are not clearly categorized into some classes. 
The generator is trained to fool the discriminator by generating target features close to the two classifiers' decision boundaries. 

In summary, this paper makes the following main contributions. 
\begin{itemize}
\item We propose a novel gesture classification framework for virtual locomotion, by regarding several consecutive frames as a point cloud in which the HMD and two VIVE trackers provide three points in each frame and each point consists of 12-dimensional features. To our knowledge, \zlz{it is \textit{the first work} for in-place gestures classification context  with considering consecutive frames as a point cloud.}
\item \zlz{We take both supervised learning and unsupervised domain adaptation into account and develop an aggregated architecture 
for joint training. }
\item We create a dataset that involves 9 classes of gestures. To our knowledge, this is \textit{the first dataset} that includes the largest number of in-place gesture classes. \textit{We will release it and our source code to the community.}
\item We conduct experiments on our dataset with the proposed method. Results show that our method is able to generate high overall classification accuracy (95.0\%) with real-time latency (192ms).
\end{itemize}

\section{Related Work}

\subsection{Walking-in-Place for Virtual Locomotion}
Virtual locomotion is a technique by which the user controls the virtual avatar to travel in a 3D virtual environment. It has a significant impact on the users' sense of presence \cite{shi2019accurate}. 
Real walking is an immersive and natural approach to navigate in the virtual environment 
\cite{riecke2010we,usoh1999walking}, but this method requires the same large physical space as the virtual environment for users to move, which in most cases can not be satisfied because the virtual space can be infinite and the room space is limited \cite{10.1145/3139131.3139133,LLCMWIP}.

WIP is a natural way of virtual locomotion using leg motions while remaining in place, which makes it easy to be applied in limited physical space. 
WIP is easy to control the direction and distance of movement  \cite{10.1145/3013971.3014010,usoh1999walking,slater1995taking} and it can also reduce simulation sickness  \cite{tregillus2016vr}. 

WIP motion is usually detected by wearable sensors, such as inertial measurement units (IMUs) embedded in HMD, smartphones, and trackers, etc. 
We can get the information of leg position, linear velocity, acceleration, angular velocity etc by attaching the wearable sensors to the user's body \cite{hanson2019improving}. Tregillus \textit{et al.} \cite{tregillus2016vr} used the smartphone's inertial sensors which include a 3-axis accelerometer and 3-axis gyroscope to collect the acceleration signal from the HMD in real-time and used Zhao's  proposed pedometry algorithm \cite{zhao2010full} to implement VR-Step. Wendt \textit{et al.} \cite{bobbing} built a model that uses only the bobbing head-track position data to estimate the walking speed, step frequency and direction. Paris \textit{et al.}  \cite{10.1145/3119881.3119889} used the IMU of the smartphone for Samsung Gear VR to detect whether the user is stepping by setting threshold values of the acceleration signal. Feasel \textit{et al.}  \cite{LLCMWIP} proposed a  LLCM-WIP System to reduce the system latency by using a chest-orientation tracker and foot trackers. 
Lee \textit{et al.} \cite{sensors18} used the vertical position of the HMD to detect the jogging in place step and considered eliminating the impact of the HMD's pitch angle on the vertical position. 
Wendt \textit{et al.} \cite{GUDWIP} implemented a Gait Understanding-Driven(GUD) WIP model with a fairly low start-stop latency by attaching two 6-DoF trackers to the user's shins. With prior knowledge of biomechanics, they  implemented a state machine with manual-tuning of parameters, then walking speed was estimated by the proposed GUD WIP model from step frequency and foot position.

Most of the aforementioned methods rely greatly on empirical experience and require the experimenter to be familiar with the biomechanical characteristics of the gait in order to manually adjust the  parameters of the WIP model such as threshold values. Moreover, as suggested by  \cite{shi2019accurate}, there are inter-person and intra-person variations. The intra-person variation refers to the fact that the sensor signal of the same person may change significantly for multiple repetitions of the same experiment due to physical exertion, and the inter-person variation refers to the fact that the sensor signal of the same gesture may vary significantly for different users, due to the differences in height, weight, gender and movement habits among users. These factors lead to significant differences in the effectiveness of manually setting model parameters for different users \cite{hanson2019improving}.

\subsection{Machine Learning for Walking In Place}
Previous data-driven approaches for WIP were usually used to recognize certain patterns of motion. Slater \textit{et al.} \cite{razzaque2002redirected} used the position of the HMD as input to a simple neural network to identify whether the user is walking in place or not. Hanson \textit{et al.} \cite{hanson2019improving} utilized the linear acceleration of the x, y, z axes of the HMD as input and fed it into a convolutional neural network to predict whether the user is currently walking or standing in real-time. Their approach  improves the generalization of the WIP model and reduces the impact of inter-person variation across different users on the model. 
Lee \textit{et al.} \cite{lee2019predicting} used the position values of the x-axis and z-axis of the HMD labeled with the torso direction of the user to train a model based on the LSTM network to predict the current WIP direction of the user. 
Lee \textit{et al.} \cite{lee2020effects} attached the VIVE tracker to the user's left ankle and used the continuous x-y-z coordinates and yaw angles of the HMD and tracker labeled with the user's posture in the gait cycle to train a convolutional neural network for predicting the gait state of the user's WIP posture. \zlz{Maghoumi \textit{et al.} \cite{deepgru} proposed an end-to-end gesture recognition network consisting of an encoder network of stacked gated recurrent units, attention module and classification layers. It directly took as input a sequence of raw 3D gesture vector features collected by sensors to predict gesture labels. }

Shi \textit{et al.} \cite{shi2019accurate} attached a microchip and three pressure sensors to the insole, which were then used to collect pressure signals from users under seven different gait patterns to create a dataset. This dataset was used to train a pattern classifier for real-time classification of user's gait, which is based on the LSTM. Shi \textit{et al.} then proposed a Dual-Check Till Consensus (DCTC) method, which dynamically adjusted the segment length of the input data according to the classifier's probability, to improve the classification accuracy of the LSTM and to reduce the system latency. Their approach can achieve excellent generalization across sensors and individuals.

\subsection{Deep Learning on Point Clouds}
A point cloud is a set of unordered 3D points, which is the most direct way to represent 3D geometric information \cite{pcservy}. 
PointNet \cite{qi2017pointnet} was pioneer in directly consuming raw point cloud representation without either voxelization or projection. It obtains permutation invariance by a symmetric function. Qi \textit{et al.} further proposed PointNet++ \cite{qi2017pointnet++}, a hierarchical network that extracts the geometric features of each point's neighborhood. DGCNN \cite{wang2019dynamic} is a graph-based method which creates a dynamic graph in the feature space and uses the EdgeConv layer to learn the edge features of the graph. Guo \textit{et al.} \cite{guo2021pct} proposed  Point Cloud Transformer (PCT) to apply the traditional  Transformer \cite{vaswani2017attention} to the field of point cloud learning. They constructed an attention mechanism based on the Transformer for the learning of point cloud features. The Transformer has the order-invariant property, which is favorable for processing point clouds. Zhang \textit{et al.} proposed to filter 3D point clouds with a point-based network \cite{zhang2021}. \zlz{Vatavu \textit{et al.} \cite{gesturesaspc}  proposed to consider the two-dimensional strokes generated by gestural input as a disordered set of points, and introduced a \$P recognizer, a member of the \$-family, to recognize the input stroke gestures. 
}

\subsection{Unsupervised Domain Adaptation (UDA)  for Point Clouds}
Qin \textit{et al.}\cite{qin2019pointdan} proposed PointDAN, which aligns the local and global features of point clouds together using a Self-Adaptive (SA) node module with an adjusted receptive field. They also used an adversarial training strategy to learn and align features across domains. 
Achituve \textit{et al.} \cite{achituve2021self} applied self-supervised learning(SSL) to point cloud processing. They came up with a multi-task architecture with a multi-head network, where one head is used to train the classifier and one head is trained using the loss caused by the self-supervised Deformation Reconstruction (DefRec) task. 
SqueezeSetV2 \cite{wu2019squeezesegv2} used geodesic correlation alignment and progressive domain calibration to bridge the domain gap between synthetic and real point clouds.

\section{Method}
\begin{figure}[h] 
\centering 
\includegraphics[width=0.45\textwidth]{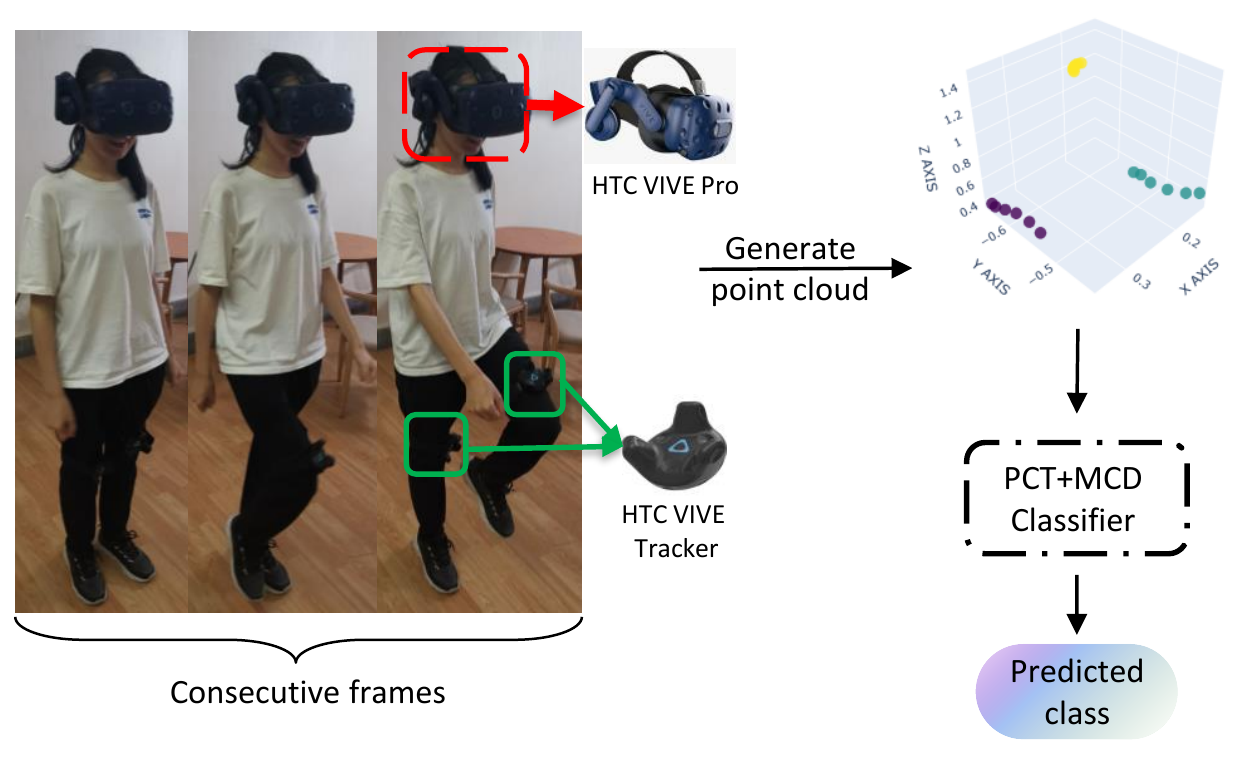} 
\caption{We \zlz{treat the data of several consecutive frames collected by the HMD and the two trackers as a point cloud sample, which is fed into the developed PCT+MCD classifier for gesture prediction in real time.} } 
\label{abstract_fig_crop} 
\end{figure}

The goal of our research is to build a model that can quickly and accurately identify the user's in-place gestures for virtual locomotion.
To this end, we first collect locomotion data using the HTC VIVE Pro HMD and two VIVE trackers attached to the user's thighs, in which each device provides a point at each frame. Each point involves the 3D position coordinates, velocity, rotation and angular velocity of the user's head or legs. We treat the data of several consecutive frames as a point cloud sample and build a point cloud dataset by labeling all samples. To consider the inter-person variations, we design a point-based classification network by fusing unsupervised domain adaptation with a point cloud backbone for joint supervised and unsupervised training. \zlz{A general overview of our approach is shown in Figure \ref{abstract_fig_crop}}.

\subsection{Dataset}
\label{sec:dataset}

\subsubsection{Participants}
To collect the dataset for training the in-place gesture classification model, we recruit 14 volunteers, all of whom are students from our university, with a mean age of 24.3 and a standard deviation of 2.1. \zlz{In them, 5 are female and 9 are male.}
Before data collection begins, we ask the volunteers about their familiarity with virtual reality using a questionnaire from ~\cite{shi2019accurate}. The average  familiarity score is 2.9 and the standard deviation is 0.87. According to ~\cite{shi2019accurate}, these numbers represent a moderate level of familiarity with virtual reality for this group of participants. Each participant is informed of the methodology and purpose of the study before the experiments.

\subsubsection{Selection of Gestures}
\label{sec:selecofgest}
\begin{figure*}[h] 
\centering 
    \subfigure[]{
        \begin{minipage}[t]{0.11\linewidth}
            \centering
            \includegraphics[width=\linewidth]{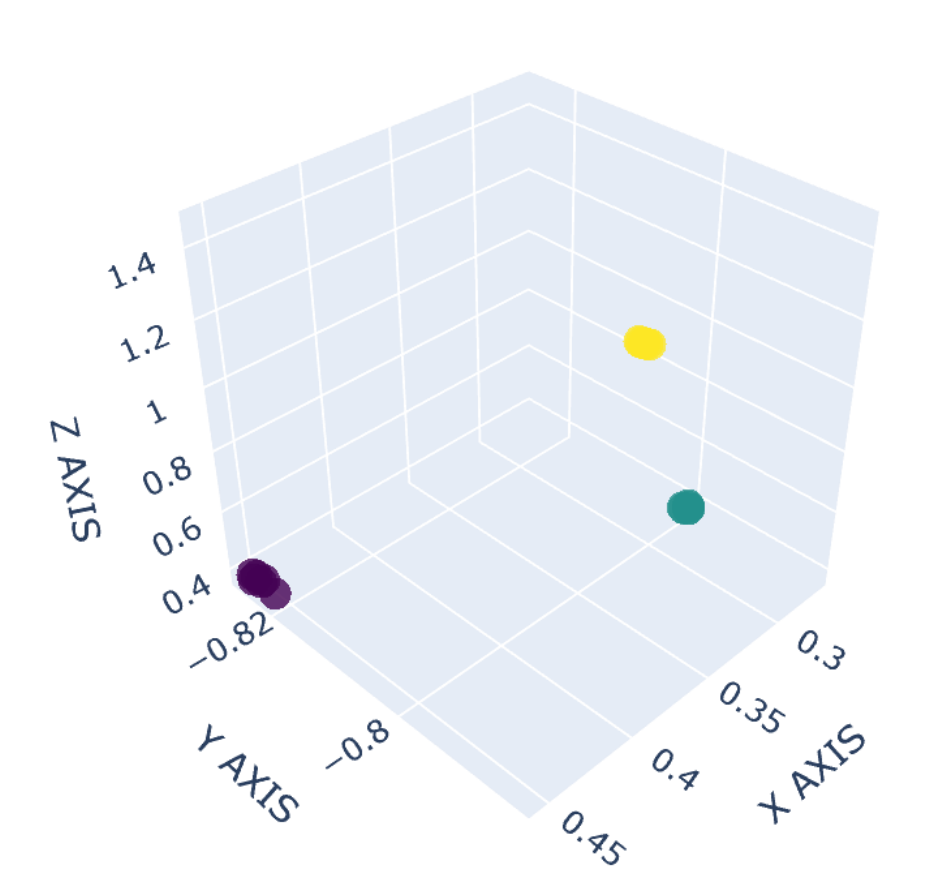}
        \end{minipage}%
    }%
    \subfigure[]{
        \begin{minipage}[t]{0.11\linewidth}
            \centering
            \includegraphics[width=\linewidth]{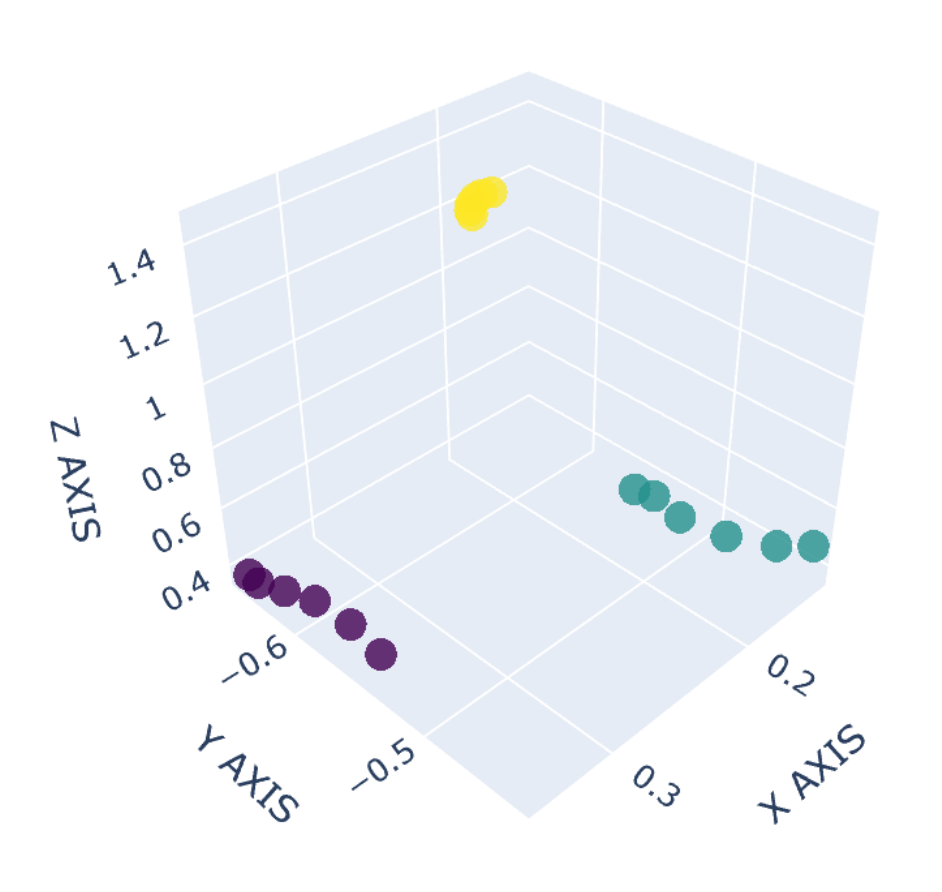}
        \end{minipage}%
    }%
    \subfigure[]{
        \begin{minipage}[t]{0.11\linewidth}
            \centering
            \includegraphics[width=\linewidth]{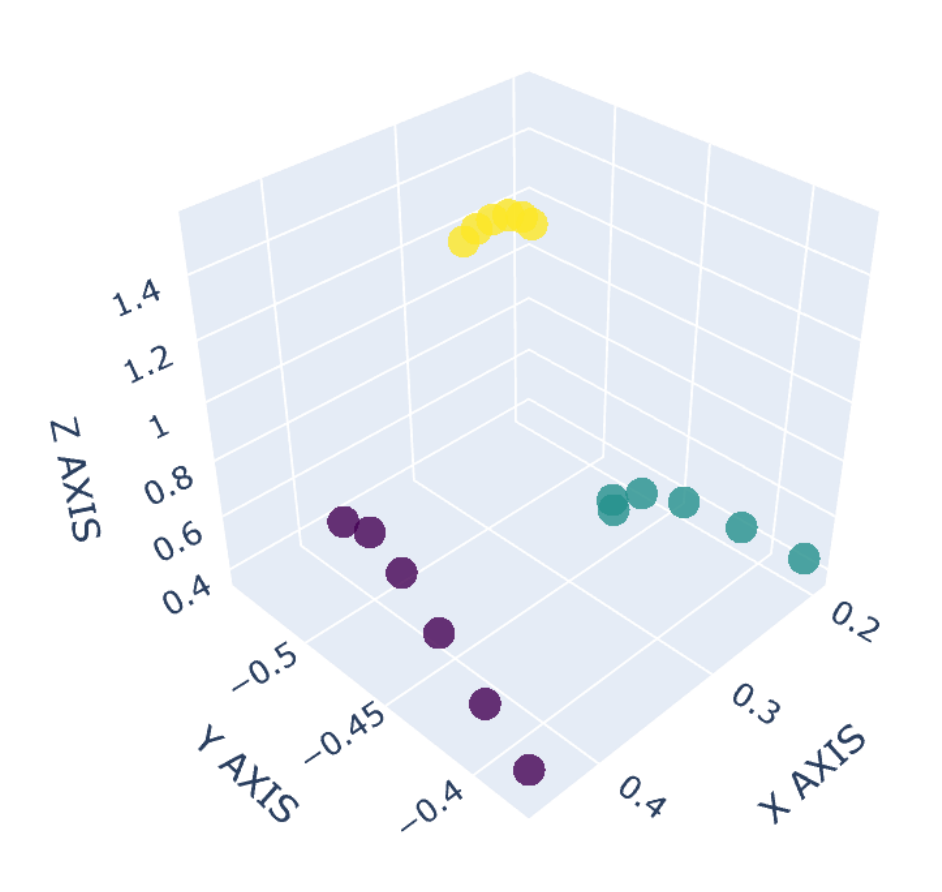}
        \end{minipage}%
    }%
    \subfigure[]{
        \begin{minipage}[t]{0.11\linewidth}
            \centering
            \includegraphics[width=\linewidth]{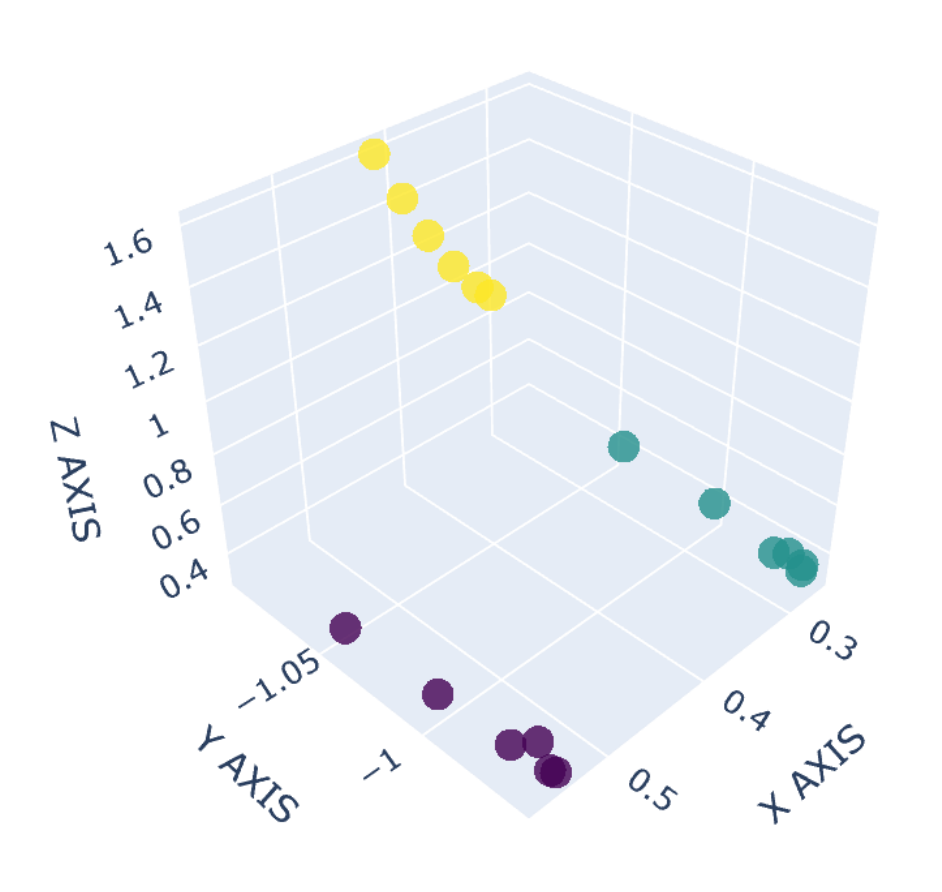}
        \end{minipage}%
    }%
    \subfigure[]{
        \begin{minipage}[t]{0.11\linewidth}
            \centering
            \includegraphics[width=\linewidth]{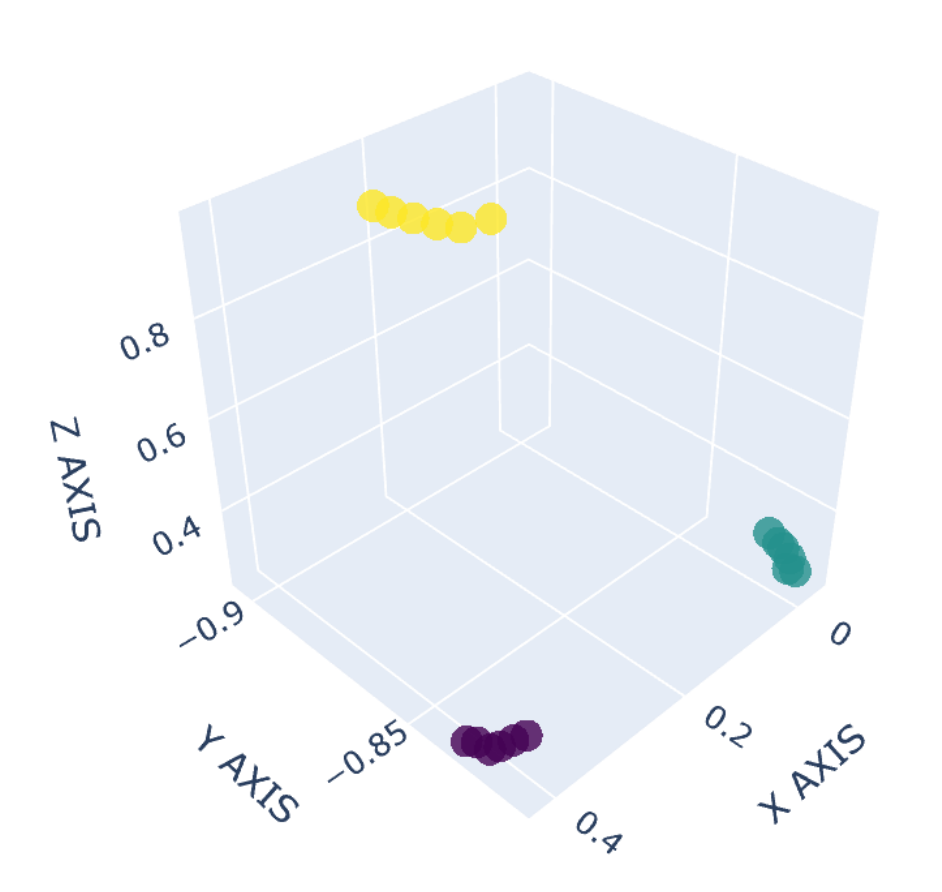}
        \end{minipage}%
    }%
    \subfigure[]{
        \begin{minipage}[t]{0.11\linewidth}
            \centering
            \includegraphics[width=\linewidth]{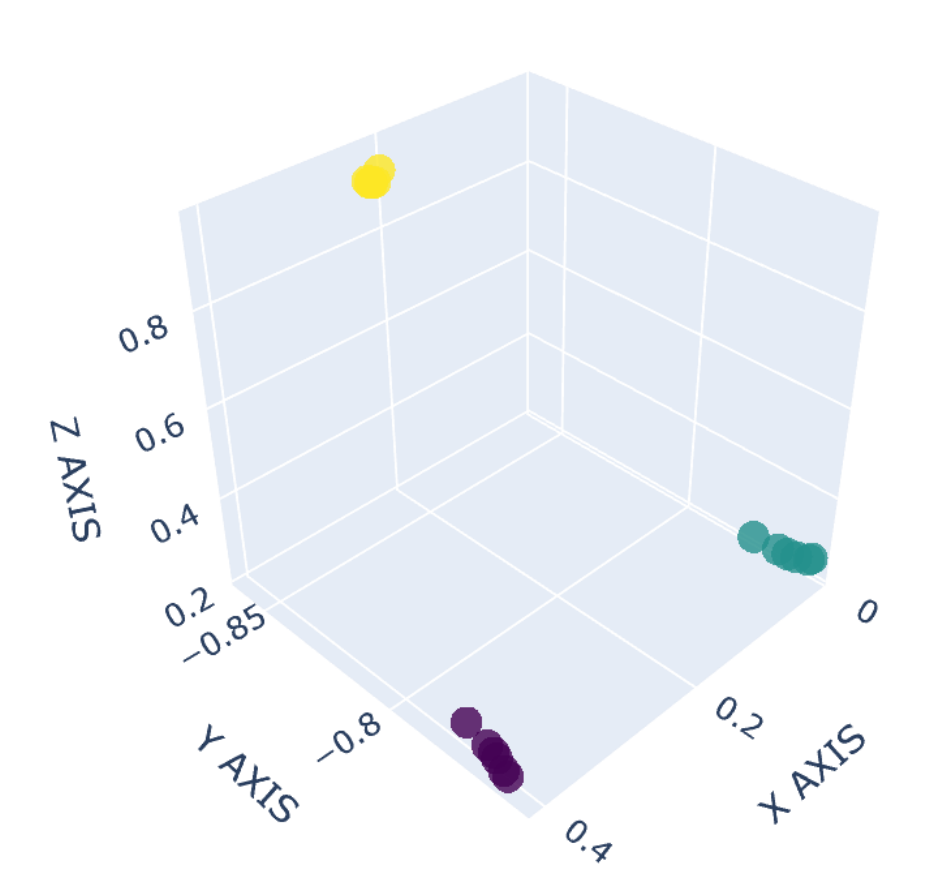}
        \end{minipage}%
    }%
    \subfigure[]{
        \begin{minipage}[t]{0.11\linewidth}
            \centering
            \includegraphics[width=\linewidth]{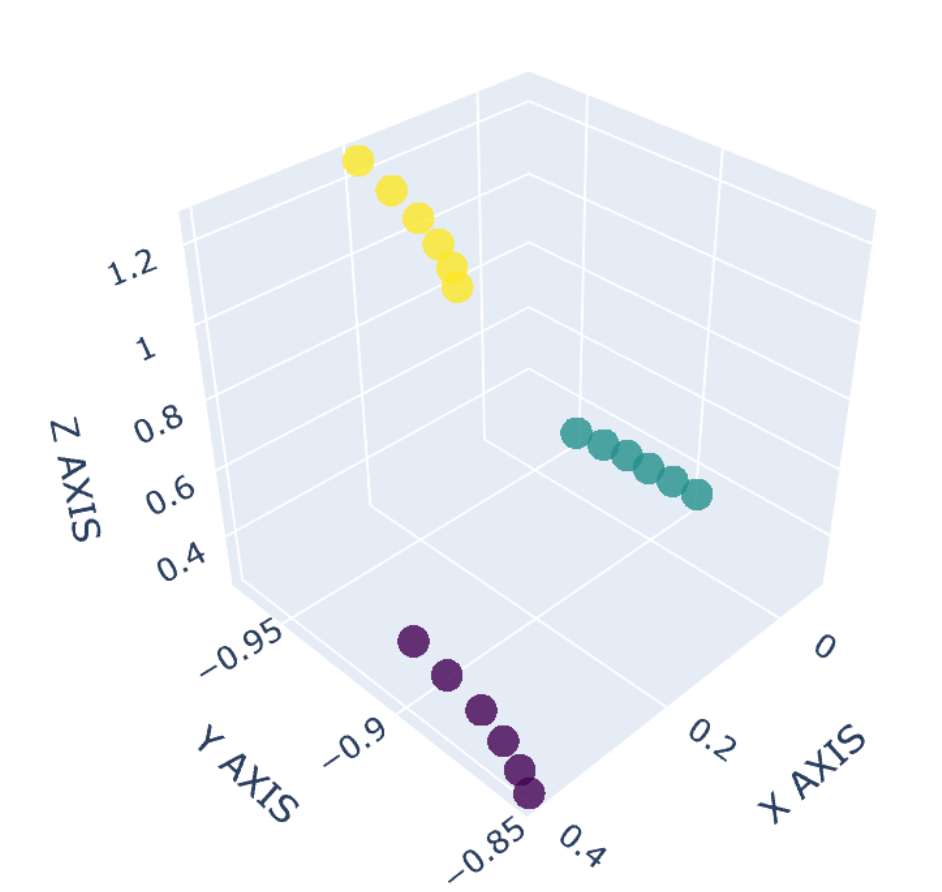}
        \end{minipage}%
    }%
    \subfigure[]{
        \begin{minipage}[t]{0.11\linewidth}
            \centering
            \includegraphics[width=\linewidth]{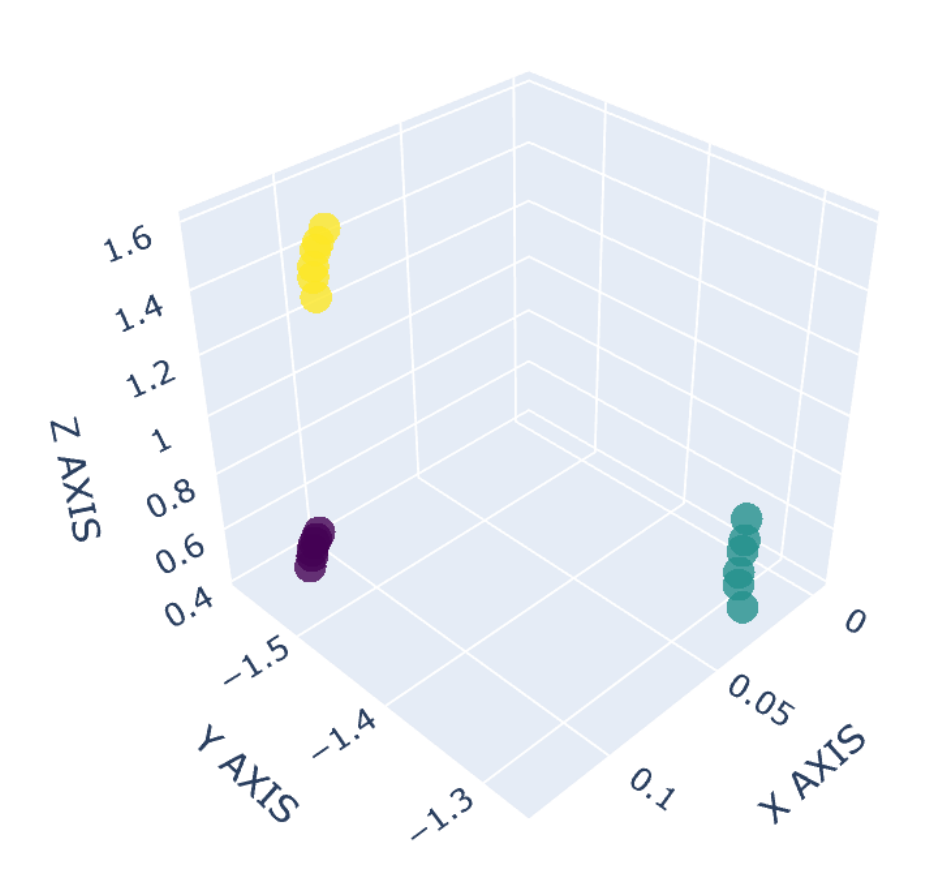}
        \end{minipage}%
    }%
    \subfigure[]{
        \begin{minipage}[t]{0.11\linewidth}
            \centering
            \includegraphics[width=\linewidth]{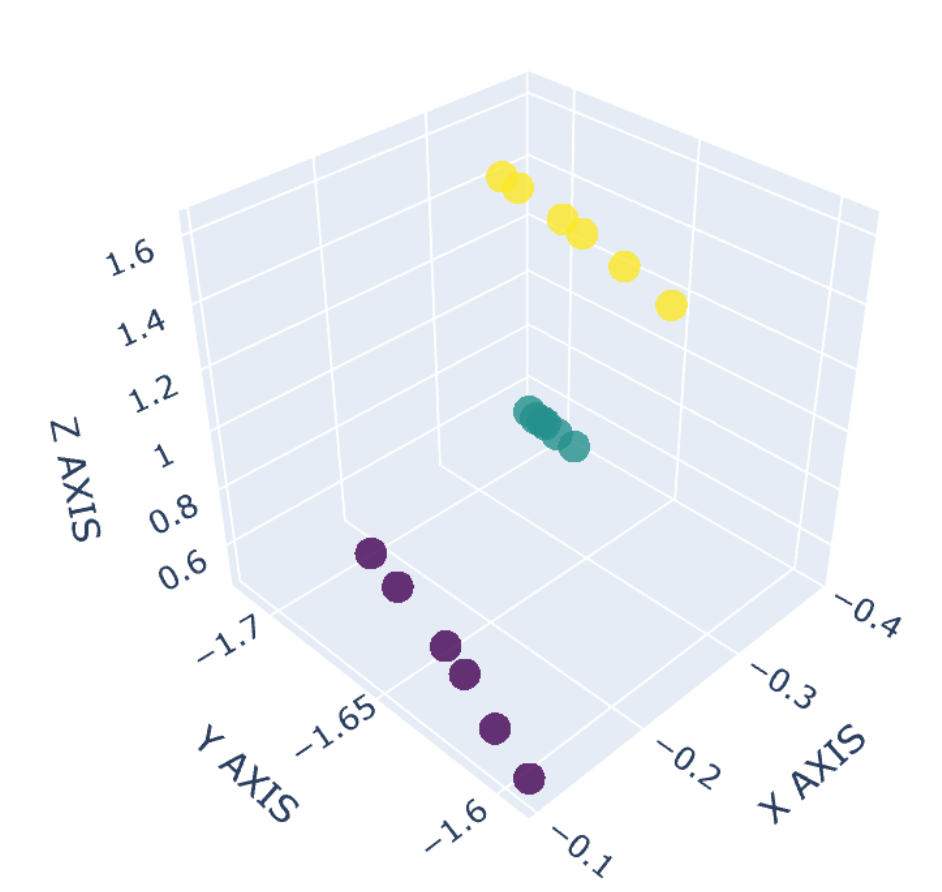}
        \end{minipage}%
    }%

\caption{Point cloud data visualization of 9 types of gestures for virtual locomotion (after normalization). (a): standing, (b): walking in place, (c): jogging in place, (d): jumping, (e): squatting down, (f): keeping squatting, (g): squatting up, (h): stepping forward, (i): stepping backward.
Different colors denote different tracking sources. 
Yellow: HMD, Green: right VIVE tracker, Purple: left VIVE tracker. } 
\label{9pc} 
\end{figure*}

There are many common forms of body movement in virtual reality games or applications~\cite{10.1145/3173574.3173908, 10.1145/3027063.3053203,hayashi2019redirected,10.1145/3313831.3376821,10.1145/3313831.3376243}. Referring to previous work \cite{tregillus2016vr,5444812,4476598}, we choose walking in place as the basic way to navigate through the virtual environment while keeping users in a limited horizontal space. In order to allow the user to control the speed \cite{hanson2019improving} and move quickly in the virtual environment, we also consider the ``jog-in-place'' motion \cite{sensors18}. Furthermore, we use real steps forward or backward as triggers to control whether the direction of walking or jogging in place is forward or backward. In the existing work \cite{10.1145/3313831.3376821}, jumping is a form of movement with high presence, creativity and immersion, so we also add jumping as a form of locomotion in our dataset. VR exergames have been shown to be effective in engaging and motivating sedentary users to exercise \cite{shaw2015development}, and squatting is a common 
gesture in exergames \cite{yoo2017evaluating}. Therefore, we add three labels to our dataset to identify squat-related gestures: squatting down, keeping squatting, and squatting up. Finally, with considering standing, our dataset involves a total of 9 labels: walking in place, jogging in place, jumping, stepping forward, stepping backward, squatting down, keeping squatting, squatting up and standing. To our knowledge, our dataset has the largest number of classes for virtual locomotion.  The normalized point cloud visualization of these 
gestures is shown in Figure \ref{9pc}, where each subplot represents a gesture and each point is provided by the HMD or VIVE tracker. Different devices are indicated by different point colors. 
(a)standing,(b)walking in place,(c)jogging in place,(d)jumping,(e)squatting down,(f)keeping squatting,(g)squatting up,(h)stepping forward,(i)stepping backward

\subsubsection{Collection Procedure}

We used the HTC VIVE Pro HMD\footnote{https://business.vive.com/us/product/vive-pro-eye-office/} and two VIVE trackers\footnote{https://www.vive.com/us/accessory/vive-tracker/} attached to the front of the participant's thighs to collect 6 DoF head and leg motion tracking data from the participants. We developed a Unity3D application to collect 3D spatial position coordinates, velocity, rotation angle and angular velocity information from the HMD and VIVE trackers at a frequency of 30 Hz. The HMD rendered a virtual scene of a beautiful mountain village with the participants positioned on the flat ground of the village.

After the experimenter helped the participants to put on the HMD and VIVE trackers, the participants followed the experimenter's instructions and performed the aforementioned 9 gestures as described in Section  \ref{sec:selecofgest}  sequentially, and each gesture lasted for about two minutes. The participants could see the current gesture name and duration timer in the HMD. Before data were formally recorded for each gesture, the experimenter instructed participants to familiarize themselves with the corresponding gesture, and this step lasted approximately one minute per gesture. The experiment can be paused at any time when the participant feels fatigued until the participant is well-rested and feels ready to continue.

We recorded the participants' data collection experiments on video, which was used as a reference to manually label the dataset.
The total experiment time for each participant was approximately 40 minutes.

\subsection{Gesture Classification}
\subsubsection{Data pre-processing}
\label{sec:datapreprocessing}
\zlz{
We propose to transform the collected data into a point cloud dataset for in-place gesture recognition. }
Since we collect data from the HMD and the two VIVE trackers at 30Hz, we can acquire 3 points at each sampling frame, with each point containing its 3D position coordinates, velocity, rotation angle and angular velocity, i.e., a total of \(3 \times 4 = 12\) dimensional features.

We use the sliding window method as mentioned in \cite{ordonez2016deep} to segment the whole dataset into point clouds. The fixed window size is 180ms, which means 6 consecutive samples at a frequency of 30Hz, and the window step size is 90ms. Each point cloud contains \(6 \times 3 = 18\)  points, where 3 refers to three sensors and the shape of the point cloud is \( (18, 12) \).
\begin{figure}[h] 
\centering 
\includegraphics[width=0.40\textwidth]{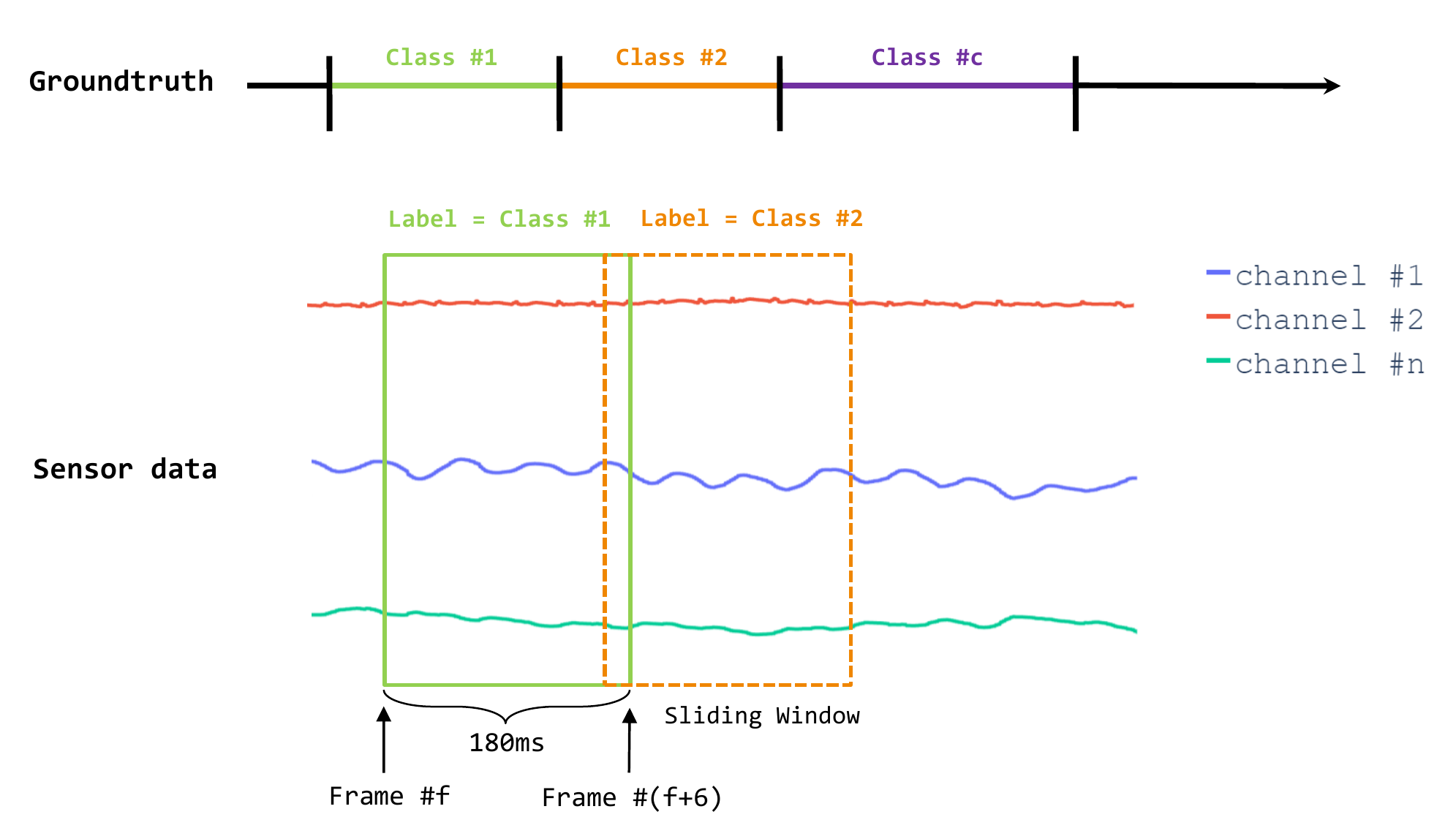} 
\caption{Splitting the data into point cloud samples with labels through the sliding window method. } 
\label{splitdataset} 
\end{figure}
The label of the point cloud indicates the ground truth of the pose contained in the corresponding window. Suppose there are \(C\) classes of gestures in the window, we choose the class that appears the most number of times in this window as the label of the window, as illustrated in Figure \ref{splitdataset}. 
The number of point cloud samples in the collected dataset is presented in Table \ref{tab:numsamples}. 

\begin{table}[tb]
  \caption{Number of point cloud samples for each gesture in our collected dataset. }
  \label{tab:numsamples}
  \scriptsize%
	\centering%
  \begin{tabu}{%
	c%
	*{2}{c}%
	}
  \toprule
  
   {Gestures} &   {Number of samples}    \\
  \midrule
  standing & 31790 \\   
  walking in place & 18409  \\
  jogging in place & 16857 \\
  jumping & 12200    \\
  squatting down & 5205 \\
  keeping squatting  & 2278 \\
  squatting up & 4688   \\
  stepping forward & 4522  \\
  stepping backward & 4457 \\
  
  \midrule
  \textbf{total} & \textbf{100406}  \\
  \bottomrule
  \end{tabu}%
\end{table}

We apply data augmentation methods for point clouds as mentioned in \cite{um2017data}, including adding Gaussian noise to jitter point clouds,  randomly rotating point clouds, shifting and scaling point clouds. Then we normalize the point clouds into \( [0,1] \) using the min-max normalization \cite{ordonez2016deep}. Data augmentation and normalization are applied to all feature channels. 

We train a model for each user, by using the current user data for unsupervised domain adaptation (target domain) and all other users' data for supervised learning (source domain). We divide the entire dataset into two parts, the labeled source domain \(\left \{X_{s}, Y_{s}\right\}\) and the unlabeled target domain \({X_{t}}\), and fuse unsupervised domain adaptation in image learning \cite{long2016unsupervised} with a point cloud backbone to train an end-to-end classifier on the target domain. Inspired by the Leave-One-Subject-Out Cross-Validation method \cite{gholamiangonabadi2020deep}, given a specific user, we use the unlabeled data of this user in the dataset as the target domain, then we remove this user's data from the dataset and use the remaining labeled data as the source domain. 
We select all the data in the source domain as the source training set. 
For the data in the target domain, we select 80\% of the data as the target training set (unsupervised) and the remaining 20\% of the data as the target test set. 
Both source training set and target training set are used to train an end-to-end gesture classification model. The target test set is used to evaluate the performance of the model. Separating the training and testing data at the individual level allows evaluating the generalization performance of the model for users without labeled data \cite{shi2019accurate}.


\subsubsection{Network Architecture}
\begin{figure*}[h] 
\centering 
\includegraphics[scale=0.6]{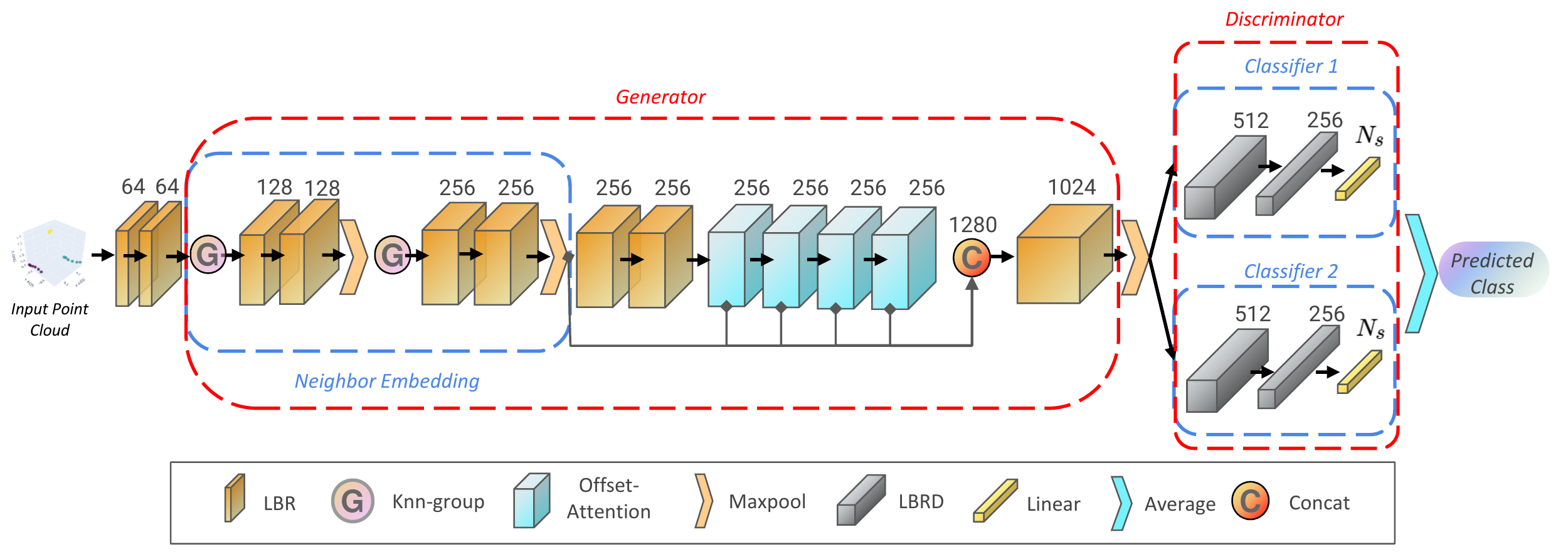} 
\caption{Our architecture (PCT+MCD). The number above each module denotes its output channel number. The LBR module represents the sequential Linear, BatchNorm and ReLU layers. The LBRD module represents the sequential LBR and Dropout layers. Knn-group module groups the point feature data into $n$ groups with k-nearest neighbors. 
\textit{In each epoch}, we first train the generator and the two classifiers with the labeled source data to minimize the cross-entropy classification loss.
Then we train the classifiers to maximize the discrepancy loss, which is defined as the difference between the outputs of the two classifiers.
Finally we train the generator to minimize the discrepancy loss. The last two steps use only unlabeled target samples.
} 
\label{netarch} 
\end{figure*}

We apply point cloud transformer (PCT)  \cite{guo2021pct} as the backbone and fuse unsupervised domain adaptation in image learning (MCD - Maximum Classifier Discrepancy) \cite{saito2018maximum}. Our model is dubbed as PCT+MCD. PCT is a  Transformer-based model for learning point clouds, achieving state-of-the-art results for point cloud classification and semantic segmentation tasks. PCT uses the neighborhood embedding module to capture the local geometric features of the point clouds and the optimized offset-attention module to learn the global features. PCT employs the neighbor embedding module combined with 4 offset-attention modules and Linear, BatchNorm, and ReLU layers as encoder, and uses three Linear layers with BatchNorm and ReLU layers as a decoder, for classification tasks. 
Inspired by the novel adversarial training strategy proposed by MCD, we use the encoder from PCT as the feature generator, and two task-specific classifiers as the discriminator which take features from the generator as input.
The two classifiers are trained to correctly classify samples in the source domain while detecting target samples that are not clearly categorized into some classes. The feature generator is trained to fool the two classifiers by generating target features close to the two classifiers’ decision boundary. After fusing them, we achieve the network architecture in our work,  as shown in Figure \ref{netarch}.

We use the encoder from PCT as the feature generator \(G\) to extract the global features of the point clouds, and we adopt two classifiers \(F_1\) and \(F_2\) consisting mainly of multiple linear layers as the discriminator. 
\(\mathbf{x}_s\) is the point cloud in the labeled source domain \(\left\{X_{s}, Y_{s}\right\}\), and \(\mathbf{x}_t\) is the point cloud in the unlabeled target domain \(\{X_{t}\}\). Both \(\mathbf{x}_s\)  and \(\mathbf{x}_t\)  contain  \(N\) points, each point with a \(d\)-dimensional feature (here $d=12$). We first feed \(\mathbf{x}_s\) and \(\mathbf{x}_t\) to \(G\) to extract the global feature \(\mathbf{f}_i=\max \operatorname{pooling}\left(G(\mathbf{x}_{i}))  \right)\), where \(\mathbf{x}_{i}\) represents $i$-th sample of  \(\mathbf{x}_s\) and \(\mathbf{x}_t\),  \(\mathbf{f}_i \in \mathbb{R}^{d}\) denotes the global features of \(\mathbf{x}_{i}\). Then we feed \(\mathbf{f}_i\) to classifiers \(F_1\) and \(F_2\) to obtain \(K\)-dimensional probabilistic outputs \(p_{1}(\mathbf{y}_i \mid \mathbf{x}_i)\) and \(p_{2}(\mathbf{y}_i \mid \mathbf{x}_i)\) which classify \(\mathbf{f}_i\) into \(K\) classes (9 kinds of gestures), respectively.

Inspired by \cite{saito2018maximum}, our total loss function consists of a classification loss and a discrepancy loss.
We adopt the cross-entropy loss function as the classification loss, which is used to train two classifiers \(F_1\) and \(F_2\)  and generator \(G\) to classify samples from the source domain  \(\left\{X_{s}, Y_{s}\right\}\):
\[\mathcal{L}\left(X_{s}, Y_{s}\right)=-\mathbb{E}_{\left(\mathbf{x}_{\mathbf{s}}, y_{s}\right) \sim\left(X_{s}, Y_{s}\right)} \sum_{k=1}^{K} \mathbf{1}_{\left[k=y_{s}\right]} \log p\left(\mathbf{y} \mid \mathbf{x}_{s}\right)\]

The discrepancy loss is defined by the \(l_1\) distance of the outputs of the two classifiers  \(F_1\) and \(F_2\) :
\[L_{d i s}\left(\mathbf{x}_{t}\right)=\mathbb{E}_{\mathbf{x}_{t} \sim X_{t}}\left[\left|p_{1}\left(\mathbf{y} \mid \mathbf{x}_{t}\right)-p_{2}\left(\mathbf{y} \mid \mathbf{x}_{t}\right)\right|\right]\]

\subsubsection{Training}
As suggested by \cite{saito2018maximum}, we also implement three steps in each epoch to realize the end-to-end training in an adversarial manner.

\textbf{Step 1.} We train both classifiers and generator simultaneously by minimizing the classification loss  \(\mathcal{L}\left(X_{s}, Y_{s}\right)\), so as to make the network correctly classify the point clouds in the source domain. 
\[\min _{G, F_{1}, F_{2}} \mathcal{L}\left(X_{s}, Y_{s}\right)\]

\textbf{Step 2.} We fix the parameters of the generator \(G\) and train two classifiers \(F_1 \) and \(F_2\) by maximizing the discrepancy loss to better detect the target samples. Classification loss is also added empirically to update the model more stably.
\[\min _{F_{1}, F_{2}} \mathcal{L}\left(X_{s}, Y_{s}\right)-\mathcal{L}_{\mathrm{adv}}\left(X_{t}\right)\]

\textbf{Step 3.} We fix the parameters of the two classifiers, and train the generator \(G\) by minimizing the discrepancy loss.
\[\min _{G} \mathcal{L}_{\mathrm{adv}}\left(X_{t}\right)\]

\section{Discussion}
\subsection{Model Evaluation}
We implement our method on PyTorch. We choose Adam \cite{kingma2017adam} as the optimizer and train the model on an Nvidia TITAN RTX GPU. The learning rate is set to 0.001 and the weight decay is assigned to 0.0001. The batch size is set to 64. Similar to  \cite{subasi2018iot}, we train an individual model for each selected user in the dataset. Each model is trained for 250 epochs. Note that for a given user, we first remove his/her data from the dataset and use  his/her unlabeled data as target and the remaining labeled data as source. 

We selected 10 subjects from the dataset marked as S1 to S10. 
4 subjects in the dataset were not selected because of missing data for the squatting gesture due to a systematic error during data collection.
For these subjects, we evaluated the classification results after training the individual models separately, including the classification accuracy of each class, the mean class accuracy (the average of each accuracy per class), and the overall accuracy (the number of correctly predicted samples divided by the number of total samples to predict). The classification results obtained for each selected subject are listed in Table \ref{tab:individual_class_result}. The results show that the average classification accuracy of keeping squatting is the lowest (85.9\%), which is caused by the relatively small amount of data of keeping squatting. During the data collection phase, subjects tended to get up immediately after squatting down, resulting in a relatively short period to maintain the keeping squatting posture. Except for keeping squatting, the average classification accuracy of all other classes is greater than 92.5\%. 
Jogging has the highest mean classification accuracy of 97.7\%. 
Note that a window size of only 180ms enables an accuracy of 94.4\% with 12ms model inference time (192ms in total),  which satisfies the requirement of real-time applications.

\begin{figure}[h] 
\centering 
\includegraphics[scale=0.3]{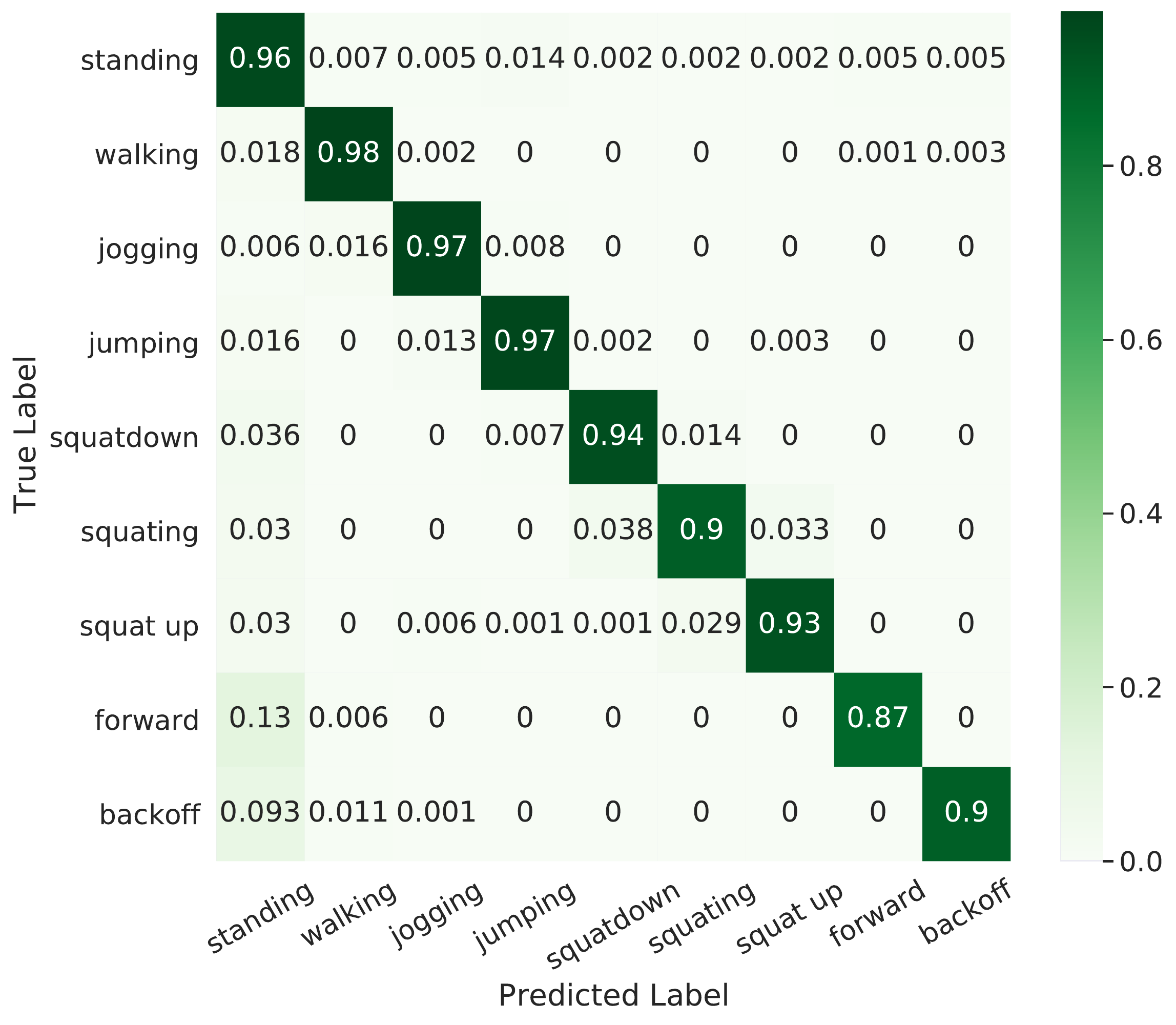}
\caption{The average confusion matrix normalized over the true (rows) condition for all the 10 selected classification models.
The horizontal axis is for the predicted class and the vertical axis is for the true class.
The diagonal elements represent the recall values. 
All other off-diagonal elements along a row are wrong predictions.
The more the correctness of a class, the green hue it has in a cell of the confusion matrix.
} 
\label{confusion_matrix} 
\end{figure}


The average confusion matrix normalized over the true (rows) condition for all the 10 selected classification models is shown in Figure \ref{confusion_matrix}. The figure shows that stepping forward achieves the lowest recall of 87\%, and stepping backward is the second-lowest recall of 90\%, and these two postures are easily misclassified as standing. 
This is probably because users are allowed to do some relaxing actions such as turning in place or resting when standing to better observe the virtual environment. 
These unintentional relaxed turns are sort of similar to forward movements or backward movements, which leads to relatively low recall of these two gestures. The recall of squatting is also relatively low (90\%) due to the relatively small sample number in the dataset.

\begin{figure}[h] 
\centering 
\includegraphics[scale=0.35]{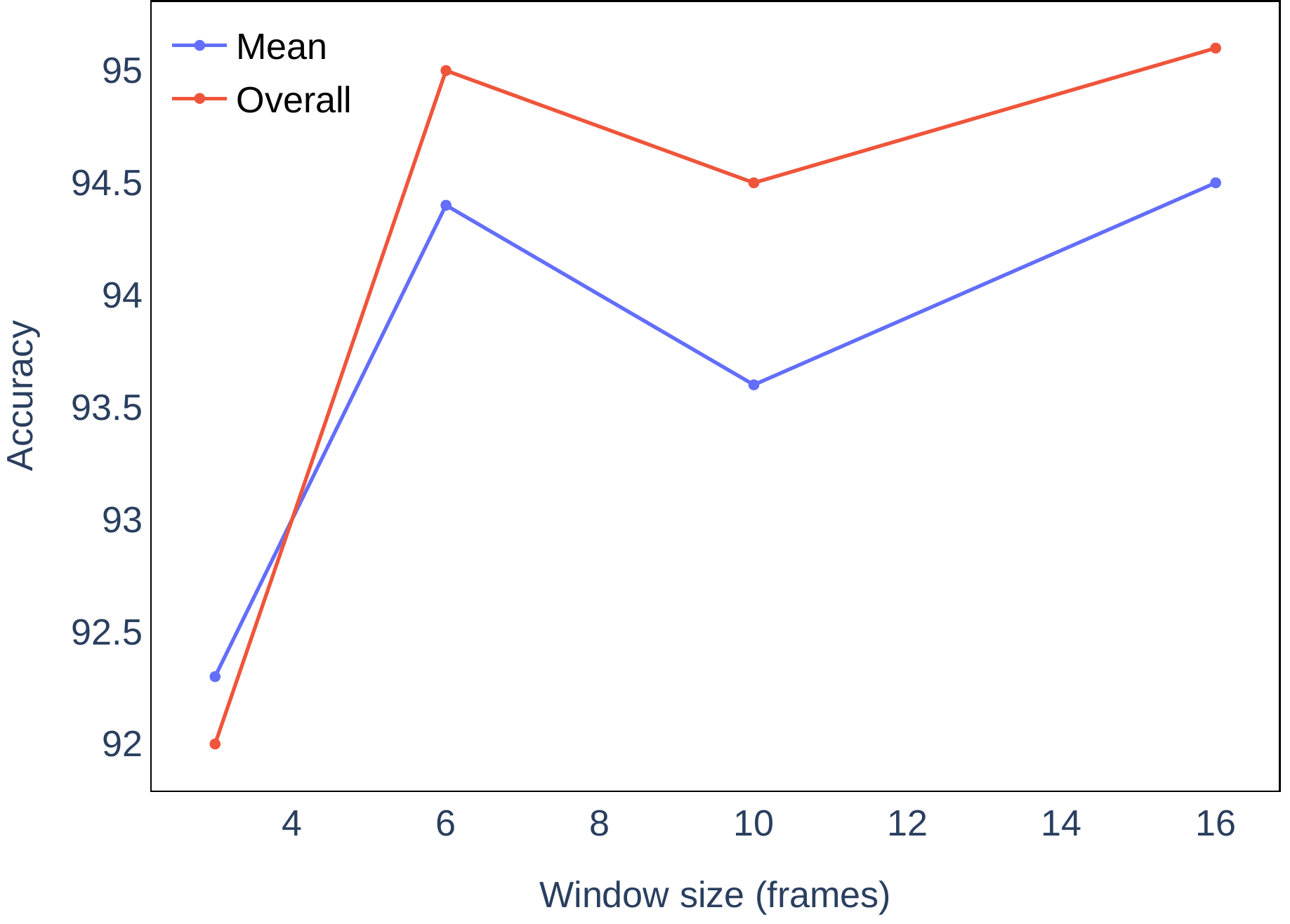} 
\caption{ \zlz{Mean class accuracy and overall accuracy with different sliding window sizes (number of consecutive frames).} } 
\label{window_size} 
\end{figure}

\zlz{The results of the model (mean class accuracy and overall accuracy) for different sliding window sizes (number of consecutive frames) are shown in Figure \ref{window_size}. The sequence duration time for generating a sample with a window size of 3 is 90ms, but the corresponding model has the lowest mean class accuracy and overall accuracy, which are only 92.3\% and 92.0\%, respectively. This is probably because a sample contains only 9 points which represent little spatial information in making gesture classification. 
When the window size is 6, the mean class accuracy and overall accuracy reach  94.4\% and 95.0\%, respectively, inducing a moderate sequence duration time (180ms). When the window size is increased to 10 and 16, we do not observe any remarkable improvement in the classification accuracy. Therefore, we choose a window size of 6 in this work.
}

\zlz{For latency, we experimentally examine the inference time of PointNet, PointNet++, DGCNN, and PCT+MCD networks on the same sample, and the results show that they all consume around 10ms. Shi \textit{et al.}  \cite{shi2019accurate} also mentioned that the maximum inference time of DCTC is around 15ms. This means the networks have very similar inference time. To achieve an accuracy of about 95\%, our method requires a sequence duration of 180ms, while DCTC requires 500ms. The inference time of these networks is significantly smaller than the sequence duration time, which indicates that the delay of ours and other machine learning methods mainly arises from the sequence duration time rather than the inference time. In short, to achieve similar accuracy, our method induces a much smaller delay compared to CNN and LSTM. This is mainly because our method only requires the collection of a few data points in a sample.}

\begin{table*}[tb] 
  \caption{Gesture classification results (\%) of our method. ``Mean'': mean class accuracy. ``Overall'': overall accuracy. ``average'' row gives the average of each column. }
  \label{tab:individual_class_result}
  \scriptsize%
	\centering%

  \begin{tabu}{%
*{12}{c}%
	}
  \toprule
  
    & {Standing} &   {Walking} &  {Jogging} &   {Jumping} &   {  Squatting  down} &   {Keeping squatting} &   {Squatting up} &   {Forward}    &   {Backward}  & Mean  & Overall  \\
  \midrule

  S1 & 97.2 & 96.3  & 100.0 & 93.6 & 100.0 & 84.8 & 100.0 & 100.0 & 100.0 & 96.9 & 97.4 \\

  S2 & 91.2 & 97.8 & 100.0 & 74.5 & 100.0 & 76.9 & 100.0 & 100.0 & 94.9 & 92.8 & 93.0 \\

  S3 & 96.6 &  100.0 & 100.0 & 99.5 & 96.4 & 90.3 & 92.6  & 94.7 & 97.2 & 96.4 & 97.9\\

  S4 & 97.3 & 95.6   & 99.0  &67.6  & 90.2   & 89.2  & 96.4   & 96.1  & 100.0  & 92.4  & 94.5 \\

  S5 & 93.8 & 97.2  & 95.2 & 97.4 & 94.9  & 94.7 & 98.4  & 92.9 & 92.5 & 95.2 & 95.3 \\

  S6 & 91.8 & 95.2  & 94.4 & 100.0 & 89.2  & 87.5  & 92.8  & 91.2 & 96.7 & 93.2 & 93.6 \\

  S7 & 95.4 & 92.2  & 98.0 & 97.3 & 91.5  & 96.4  & 87.5  & 98.4 & 90.6  & 94.2 & 95.0\\

  S8 & 87.7 & 97.1  & 100.0 & 100.0 & 100.0  & 66.7  & 97.8  & 97.5 & 97.4  & 93.8 & 95.7\\

  S9 & 85.9 & 97.9  & 94.4 & 98.9 &  91.4 & 96.3  &  94.8 & 96.4 & 98.1 & 94.9 & 93.2 \\

  S10 & 87.9 & 97.7  & 96.3 & 99.6 & 100.0  & 75.8  & 98.6 & 98.0 & 93.2&94.1 & 94.8  \\
  
  \midrule
  \textbf{average} & \textbf{92.5} & \textbf{96.7} & \textbf{97.7} & \textbf{92.8} & \textbf{95.3} & \textbf{85.9} & \textbf{95.9} & \textbf{96.5} & \textbf{96.0}  & \textbf{94.4} & \textbf{95.0} \\
  \bottomrule
  \end{tabu}%
  
\end{table*}

\subsection{Comparison with Existing WIP Methods}

Hanson \textit{et al.} \cite{8797751} applied convolutional neural networks to WIP research. They classify walking or standing using a simple CNN structure and achieved 98.6\% accuracy on the test set with a latency of 680ms (a sample is composed of 40 consecutive triaxial data and data is polled every 0.017 seconds) for the Oculus Go and 92.6\% accuracy with a latency of 1020ms (60 consecutive triaxial data) for the Samsung Gear VR. 
For comparison purpose, we trained a binary classification model using only the samples of walking and standing in the dataset. 
In contrast, our method is more accurate for classifying standing and walking, with a mean class accuracy of 99.2\%, and our method significantly shortens the latency (taking only about 192ms).

LLCM-WIP \cite{LLCMWIP} achieves a start latency of 138ms and a stop latency of 96ms, which is faster than our method. 
However, LLCM-WIP requires manual adjustment of thresholds for the model,  such as the speed-offset threshold. Due to inter-person variations, the threshold of the model needs to be adjusted across users \cite{shi2019accurate} which actually consumes time. 
In comparison, our approach avoids the need for manual setting of threshold parameters and can personalize the model using only the user's unlabeled data, which is more flexible and automatic.

DCTC \cite{shi2019accurate} can classify 7 foot patterns based on the LSTM model using the pressure sensor data. 
They achieve 95\% accuracy at 500ms delay and 85.3\% accuracy at 300ms delay. 
In contrast, our method can achieve 94.4\% mean class accuracy and 95.0\% overall accuracy at 192ms latency, which has overall better performance than DCTC with considering the latency.  
Moreover, DCTC provides different general models according to the shoe size. By contrast, our method needs to collect unlabeled data for each user to train an individual model.

\subsection{Comparison with Existing Point Cloud Classification Methods}
We compare our method (PCT + MCD) with a series of point cloud classification methods, including  PointNet \cite{qi2017pointnet}, PointNet++ \cite{qi2017pointnet++}, DGCNN \cite{wang2019dynamic} and PointDan \cite{qin2019pointdan}.  PointDan also involves unsupervised domain adaptation. These experiments are all using Adam as the optimization strategy, with a learning rate of 0.001 and a batch size of 64. The cosine annealing schedule is adopted to adjust the learning rate for each epoch. The models are trained using the same 10 subjects'  posture data as shown in Table \ref{tab:individual_class_result} and the train-test splitting method as described in Section \ref{sec:datapreprocessing}.  All models are evaluated on the test set of the target domain. The comparison results are summarized in Table \ref{tab:compare_pc_methods}. 

As illustrated in Table \ref{tab:compare_pc_methods}, our method achieves the best result of  95.0\% overall accuracy and 94.4\% mean class accuracy.
Compared to PointNet, our method achieves a 3.7\% improvement in mean class accuracy and 2.0\% improvement in overall accuracy, respectively.
Compared with PointDAN which also uses the unsupervised domain adaptation, our method achieves 4.7\% gain on mean class accuracy and 3.1\% gain  on overall accuracy, respectively. 
\zlz{
We also perform an ablation experiment, eliminating MCD and using only PCT for classification, and obtains a mean class accuracy of 92.2\% and an overall accuracy of 93.7\%, which are lower than PCT+MCD. This demonstrates the necessity of MCD. }

\zlz{
To verify the significance of the improvement of our introduced PCT+MCD, we perform two-tailed t-tests for mean class accuracy of different methods.  The t-test results show that the mean class accuracy of PCT+MCD is significantly higher than that of PointNet ($t=-3.56$, $p=.002$), PointNet++ ($t=-4.31$, $p<.001$), DGCNN ($t=-3.90$, $p<.001$), PointDAN ($t=-4.49$, $p<.001$) and PCT ($t=-3.12$, $p=.006$).
}

\begin{table}[tb]
  \caption{Comparison with other point cloud classification methods. ``Mean'' and ``Overall'': mean class accuracy and overall accuracy of the 10 selected models. }
  \label{tab:compare_pc_methods}
  \scriptsize%
	\centering%
  \begin{tabu}{%
*{12}{c}%
	}
  \toprule
  
  Method  & Mean & Overall  \\
  \midrule
  PointNet\cite{qi2017pointnet} & 90.7 & 93.0   \\
  PointNet++\cite{qi2017pointnet++} & 88.5 & 91.5   \\
  DGCNN\cite{wang2019dynamic} & 91.1 & 91.7   \\
  PointDAN\cite{qin2019pointdan} & 89.7 & 91.9   \\
  PCT\cite{guo2021pct} & 92.2 & 93.7   \\
  \midrule
  PCT+MCD(ours) & \textbf{94.4} & \textbf{95.0} \\
  \bottomrule
  \end{tabu}%
\end{table}

\section{User Study}
In this section, we conducted user study using the evaluation testbed proposed by \cite{testbed} to evaluate the straight-line walking and directional control performance, agility, and user experience of our model for locomotion in the virtual environment.

\subsection{Participants}
\zlz{
15 volunteers participated in this experiment}, each of them use their individual model trained by labeled dataset and their unlabeled data to perform the virtual locomotion experiment. Note that if the participant participated in the dataset collection experiment as mentioned in Section \ref{sec:dataset}, we remove his/her data from the dataset before training. 
There are 4 female volunteers and 11 male volunteers, and their mean age is 25.6 and the standard deviation is 2.8. The mean value of their familiarity with VR was 2.8, and the standard deviation was 0.82. 
None of these volunteers suffered from severe motion sickness.

\subsection{Experimental Setup}
We utilize 1 square meter of space in the center of the room for moving in place and install two SteamVR Base Stations around the space. Participants wear HTC VIVE Pro HMD and are attached two VIVE trackers to their left and right thighs respectively.  
Participants then performed a series of movements in place as described in Section \ref{sec:selecofgest}.

We develop the virtual locomotion system in Unity3D. 
Our model is deployed to the server through the torchserve\footnote{https://pytorch.org/serve/} framework, which allows Unity clients to call the prediction model via HTTP requests. 
We sample the user's motion data at 30Hz and input samples to the model every 180ms to predict the user's gesture class.  
The user's forward direction is defined as the average value of the z-axis rotation of the two VIVE trackers. 
When a backward step is detected, we reverse the forward direction, and when a forward step is detected, we reset the forward direction. 
When detecting a user walking in place or jogging in place, we calculate the time between steps (\(t_{step}\)), which is defined as:
\[t_{step} = |t_{maxposy} - t_{minposy}|\]
where \(t_{maxposy}\) refers to the time corresponding to the last maximum peak of the vertical position of the HMD, and \(t_{minposy}\) refers to the time corresponding to the last minimum peak of the vertical position of the HMD. A large \(t_{step}\) means that the user is walking slowly and vice versa. 
Then we calculate the virtual locomotion speed of walking (\(v_{walking}\)) and jogging (\(v_{jogging}\) ) with the equation proposed by \cite{tregillus2016vr}:
\[v_{walking}=\frac{t_{\text {step }}-I_{\min }}{I_{\max }-I_{\min }} *\left(V_{\max }-V_{\min }\right)+V_{\min }\]
\[v_{jogging} = k * v_{walk}\]
where \([I_{max}, I_{min}]\) are respectively the upper and lower bounds of \(t_{step}\), and \([V_{max},V_{min}]\) are respectively the maximum and minimum values of velocity. \(k\)  is a constant greater than 1. When \(t_{step}\) is less than \(I_{min}\), we set  \(v_{walking}\) to \(V_{max}\), and when the user is detected to be standing, we set \(v_{walking}\) to 0. When a user is detected to be jumping, we apply a forward and upward force to the rigid body of the user's virtual avatar, so that the avatar jumps forward. Finally, when the user is detected to be squatting down or keeping squatting , we lower the avatar's view and collision body, and when the user is detected to be squatting up, we reset the avatar's view and collision body. 

We selected several scenarios and tasks from the testbed \cite{testbed} which are relevant to our locomotion gestures for experiments. 
The first scenario is called Straight Line Movement and consists of 4 tasks. 
Task 1 requires the user to walk towards the target spot, following the green straight line on the ground as closely as possible. Task 2 requires the user to walk through three progressively smaller circular areas in sequence. Task 3 requires the user to follow a circle, staying inside the circle as much as possible, which moves forward at a variable speed. 
Task 4 requires the user to dash through a straight corridor as fast as possible. 
The second direction control scenario is used to evaluate the directional control performance of the locomotion method and includes  5 tasks. 
First, the user walks along the folded lines marked on the ground to reach several targets, then the user must walk backward through a door while gazing at the target ahead, after that, the user adjusts the appropriate direction through a narrow and curved tunnel, without hitting the walls as much as possible. 
Next, the user jumps through a staircase and  walks across a ramp. 
Finally, the user crosses a dangerous area on the edge of the roof of a tall building and avoids falling from the roof. 
The third agility scenario is used to evaluate the agility of the locomotion method. 
This scenario requires the user to cross a corridor with blocks of various shapes moving towards the user. 
The user has to control the direction of movement and squat to avoid the oncoming blocks.

Following the instructions \cite{testbed}, we collect some objective metrics which  focus on accuracy and latency for each task during the experiment, which are classified as  the \textit{Accuracy} (AC) and the \textit{Error-proneness} (EP). 
\zlz{We recommend that readers refer to \cite{testbed} for detailed explanation of these objective metrics. } The questionnaire provided by the testbed \cite{testbed} is used to investigate user's subjective feelings, including Input sensitivity, Input responsiveness, Ease of use, Perceived errors, Appropriateness and Satisfaction, and so on. 
Then we use the SSQ tool \cite{ssq} to assess the symptoms of motion sickness caused by our experiment (if any).

For comparison, we choose the improved LLCM-WIP \cite{LLCMWIP} method embedded in the testbed \cite{testbed}. 
This method attaches two VIVE trackers to the back of the user's left and right calves. To be fair, we set its forward direction to be the average of the z-axis value of rotation of the two trackers, which is the same as our method. 
\zlz{We also trained a model of DGCNN \cite{wang2019dynamic} for comparison purpose.}
Before the experiment, participants will be assigned one of these three methods without knowing which one it is.

\subsection{Results}
\subsubsection{Error-proneness}
\begin{figure}[h] 
\centering 
\includegraphics[scale=0.12]{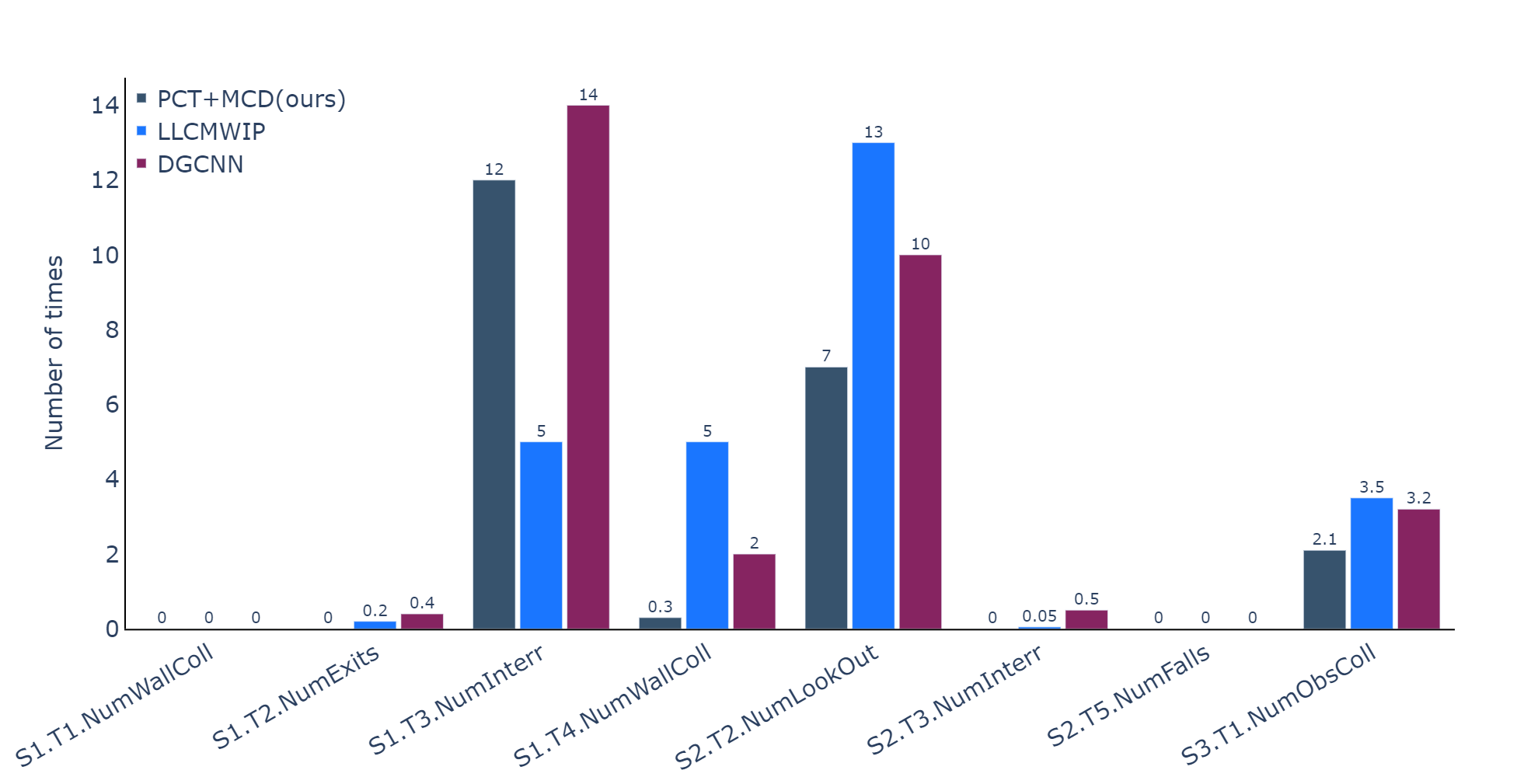}
\caption{ \zlz{Error proneness results.} } 
\label{us_error} 
\end{figure}

The results of the error proneness collected during the experiment are presented in Figure \ref{us_error}. 
The results show that our method has significantly fewer errors than the LLCM-WIP method in Task 4 of Scenario 1 and Task 2 of Scenario 2. 
This is because our method is easy to toggle the forward or backward direction of movements and has a relatively low start-stop delay. 
The LLCM-WIP method does not have a backward mechanism, and the user has to turn his body and turn his/her head backward, which leads to lower performance in the backward task. 
However, our method has more errors than LLCM-WIP in Task 3 of Scenario 1. 
In this task, the circle that the user is chasing will suddenly accelerate at a moment and the user tends to toggle from walking to jogging at this moment, so the walking speed will increase suddenly from \(V_{walking}\) to \(V_{jogging}\) instead of increasing smoothly, resulting in the user having difficulty in controlling the speed and then drifting out of the circle. 
In other tasks, both our method and the LLCM-WIP method show a small error proneness, and there is no significant difference between them. 
\zlz{ We also found a slight increase in user error-proneness when using the  DGCNN model due to a decrease in classification accuracy compared to the PCT+MCD model. }

\subsubsection{Accuracy}
\begin{figure}[!h] 
\centering 
\setlength{\abovecaptionskip}{0.cm}

    \subfigure[S1.T1.STPathDev]{
        \begin{minipage}[t]{0.3 \linewidth}
            \centering
            \includegraphics[width=\linewidth]{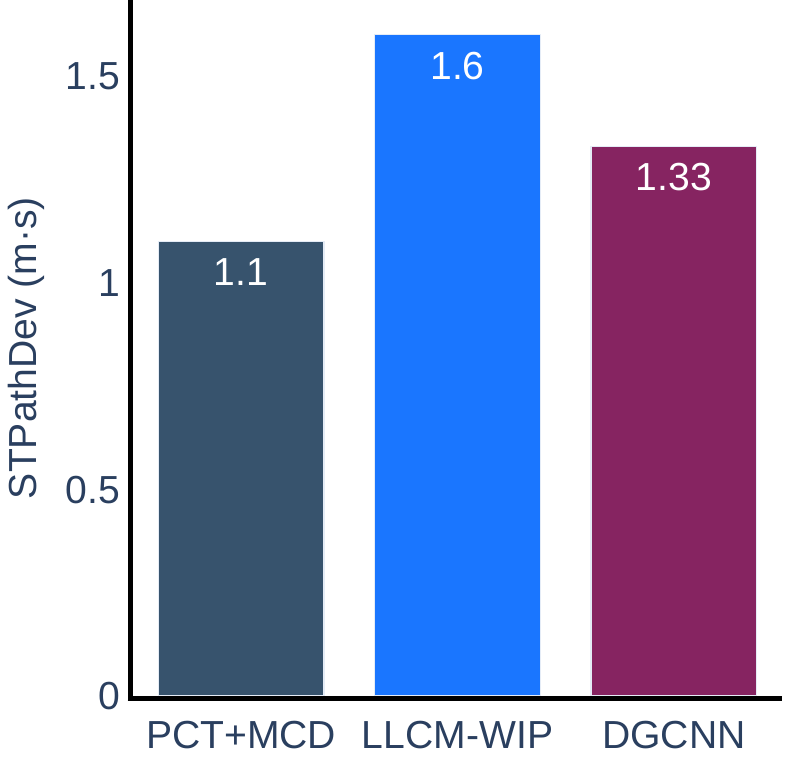}
        \end{minipage}%
    }%
    \subfigure[S1.T2.TargetDist]{
        \begin{minipage}[t]{0.3\linewidth}
            \centering
            \includegraphics[width=\linewidth]{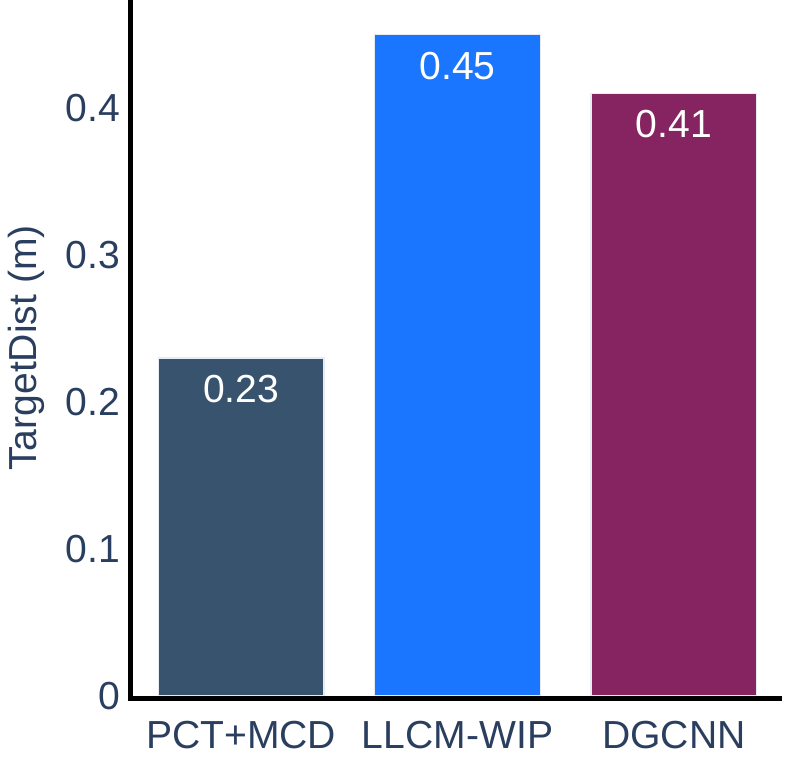}
        \end{minipage}%
    }%
    \subfigure[S1.T3.InsideTargetRate]{
        \begin{minipage}[t]{0.3\linewidth}
            \centering
            \includegraphics[width=\linewidth]{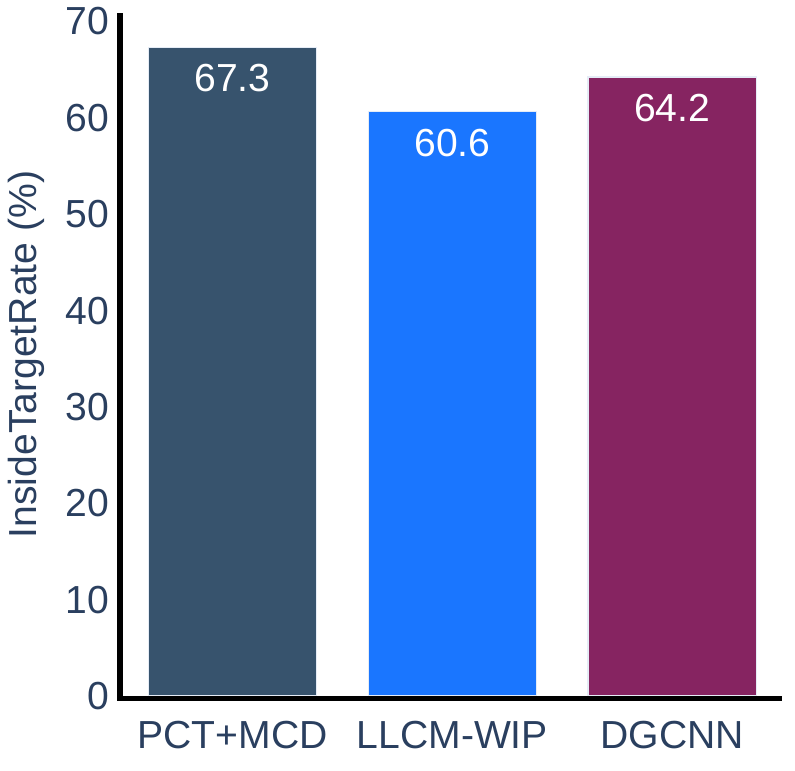}
        \end{minipage}%
    }%
    
    \subfigure[S1.T3.AvgDist]{
        \begin{minipage}[t]{0.3 \linewidth}
            \centering
            \includegraphics[width=\linewidth]{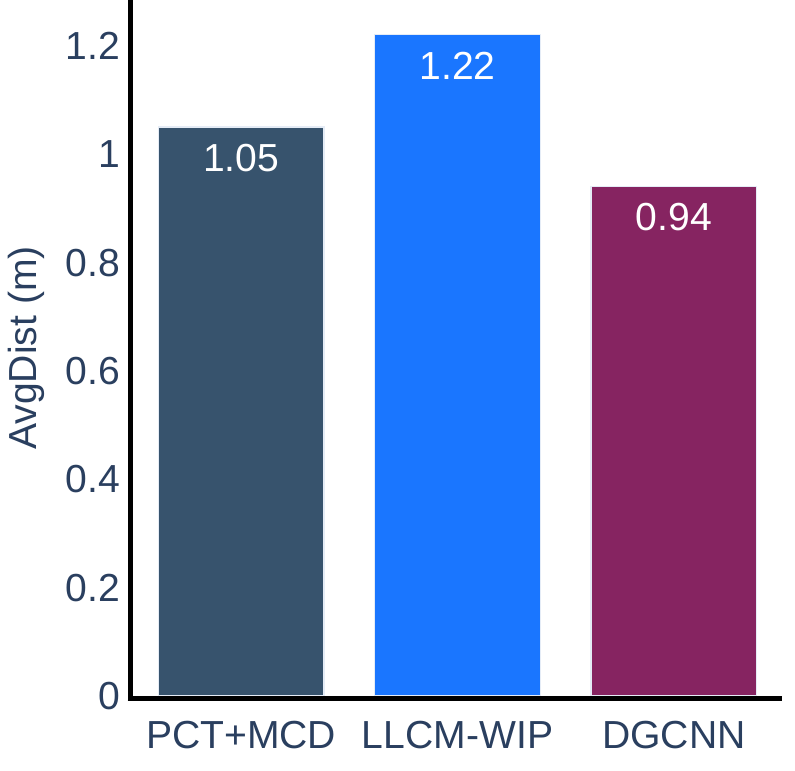}
        \end{minipage}%
    }%
    \subfigure[S2.T1.InitAngErr]{
        \begin{minipage}[t]{0.3\linewidth}
            \centering
            \includegraphics[width=\linewidth]{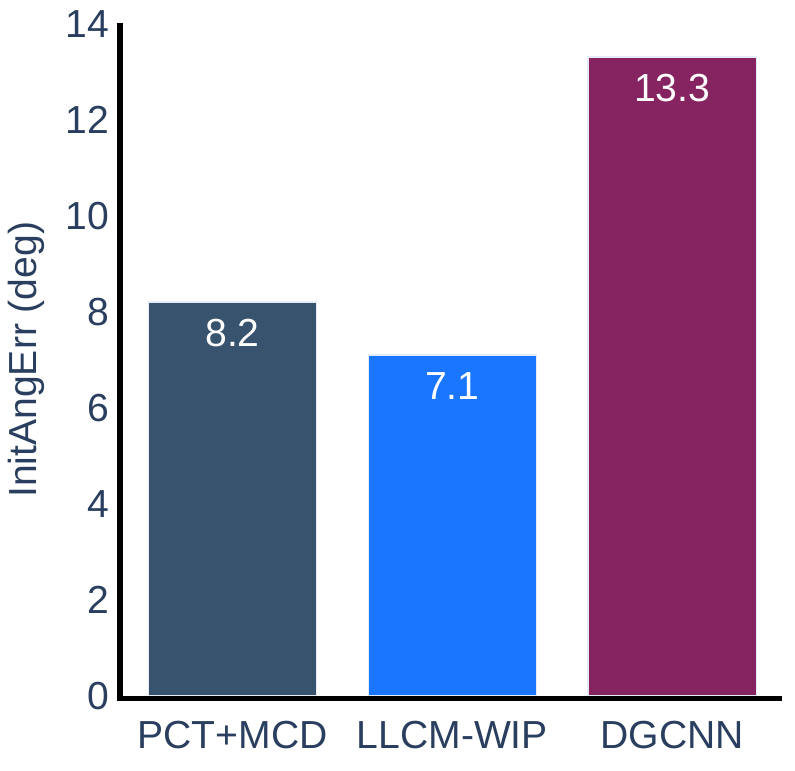}
        \end{minipage}%
    }%
    \subfigure[S2.T1.EstPathLen]{
        \begin{minipage}[t]{0.3\linewidth}
            \centering
            \includegraphics[width=\linewidth]{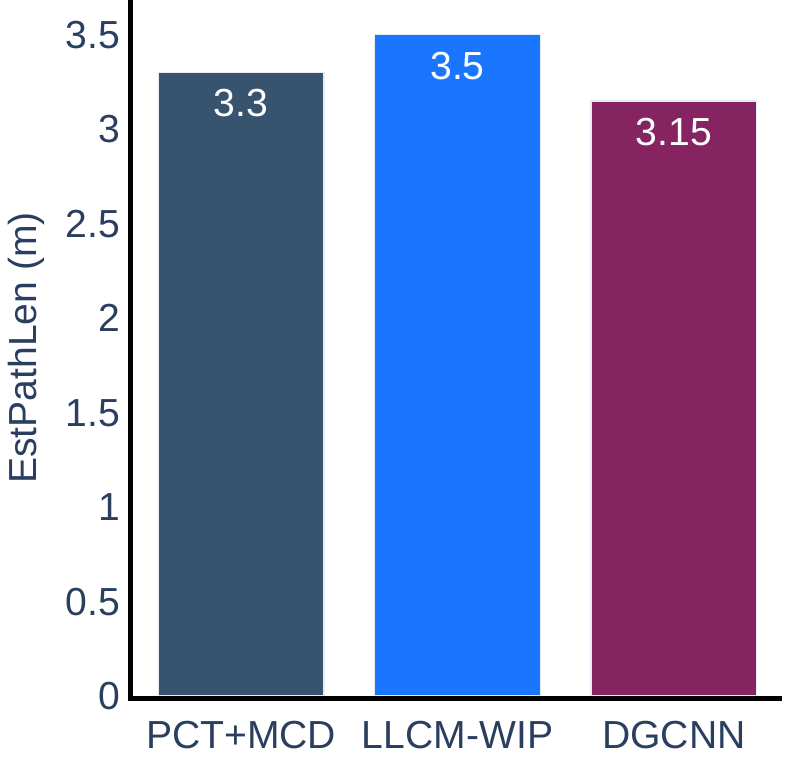}
        \end{minipage}%
    }%
    
    \subfigure[S2.T1.RecallTime]{
        \begin{minipage}[t]{0.3\linewidth}
            \centering
            \includegraphics[width=\linewidth]{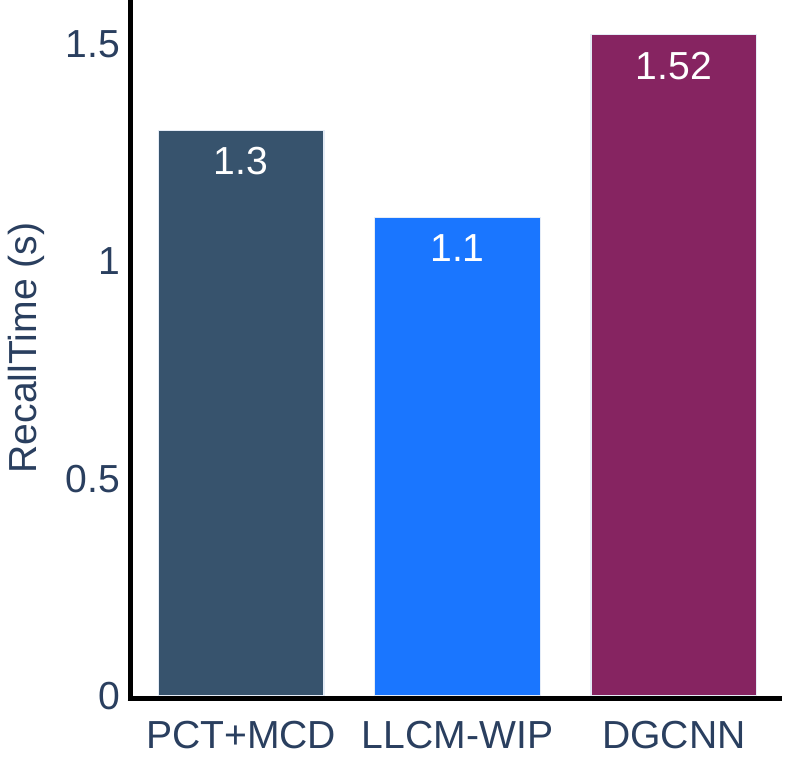}
        \end{minipage}%
    }%
    \subfigure[S2.T2.AccuracyBkw]{
        \begin{minipage}[t]{0.3\linewidth}
            \centering
            \includegraphics[width=\linewidth]{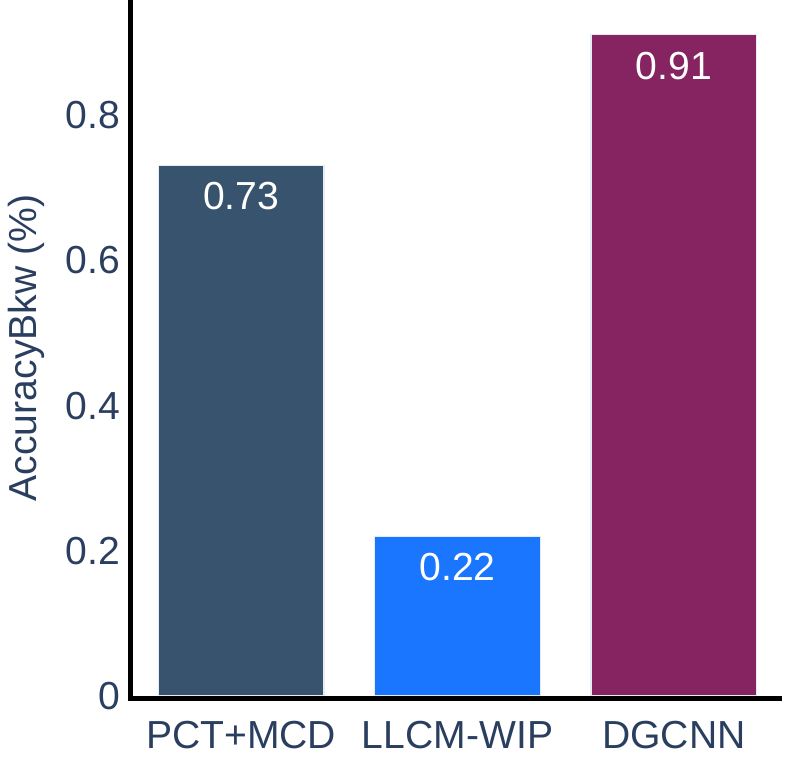}
        \end{minipage}%
    }%
    \subfigure[S2.T5.Avoidance]{
        \begin{minipage}[t]{0.3\linewidth}
            \centering
            \includegraphics[width=\linewidth]{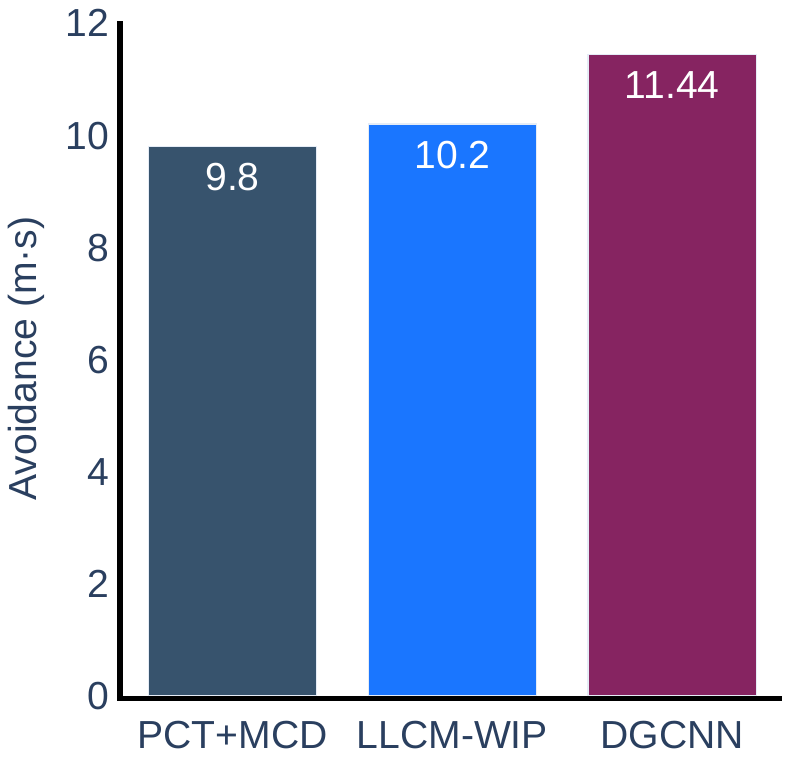}
        \end{minipage}%
    }%

\caption{ \zlz{User study accuracy results. Each subplot illustrates an objective metric.}} 
\label{us_acc_fig} 
\end{figure}

The experimental results for the accuracy metric are presented in Figure \ref{us_acc_fig}. 
We can conclude that our method is more accurate than LLCM-WIP for Task 1 and Task 2 in Scenario 1 and Task 2 in Scenario 2. When using our method, the user can walk closer to the target straight line indicated on the ground and can stop more accurately within the target range. 
The results on other tasks do not show any significant difference between our method and LLCM-WIP. 
\zlz{When compared with DGCNN, our method is more accurate on tasks of Scenario 2, but slightly less accurate than DGCNN on tasks of Scenario 1. }

\subsubsection{Subjective Evaluation}
We conducted a simulator sickness questionnaire before and after the experiment, and we found that the  experiment did not induce motion sickness symptoms.

For latency, we asked users to respectively evaluate the input responsiveness questionnaire at the end of each scenario experiment and at the end of the full experiment. 
Users are asked to rate two statements in a scale from 1 (Strongly Disagree) to 5 (Strongly Agree), with the first statement being ``The response to user input was acceptable'' and the second being ``The response time did not affect my performance''. 
\zlz{The results show that the average rating of PCT+MCD and LLCM-WIP are 4.7 (SD: 0.48) and 4.9 (SD: 0.32), respectively. We then perform the two-tailed t-test at the 5\% significance level and did not find a statistically significant difference in the rating of PCT+MCD and LLCM-WIP delays ($t = 1.095$, $p=.288$).}

\section{Conclusion}
We presented an effective classification framework for classifying 9 in-place gestures for virtual locomotion.  
We propose to treat several consecutive frames as a point cloud in which the HMD and two VIVE trackers provide three points in each frame. 
We design an end-to-end method by fusing supervised learning with unsupervised domain adaptation. 
Experiments show that our method achieves very promising outcomes, i.e., 95.0\% overall accuracy and a latency of 192ms. We also conducted user study to further verify the effectiveness of our method. 

However, the manual labeling of the dataset is time-consuming and tedious, which limits us to build a larger dataset. In future, we would like to speed up this with the aid of other techniques like pose estimation\cite{Toshev_2014_CVPR}.
\zlz{We also plan to increase the number of IMUs, and fuse deep learning methods with traditional biomechanics-based methods to improve accuracy.}

\acknowledgments{
This work was partially funded by 
2021 Science and Technology Innovation Program of Shaanxi Academy of Forestry Science (SXLK2021-0214),
Key Laboratory of Agricultural Internet of Things, Ministry of Agriculture and Rural Affairs, Yangling,Shaanxi 712100, China (2018AIOT-09). 
National Natural Science Foundation of China(61702433).
The Key Research and Development Project of Shaanxi Province (2019NY-167).
}


\bibliographystyle{abbrv-doi}

\bibliography{template}
\end{document}